\documentclass[12pt]{article}
\usepackage[top=1in, bottom=1.1in, left=1in, right=1in]{geometry}
\usepackage{setspace} 

\def\spacingset#1{\renewcommand{\baselinestretch}%
{#1}\small\normalsize} \spacingset{1}
\spacingset{1.8} 

\usepackage[utf8]{inputenc}
\usepackage[T1]{fontenc}
\usepackage{amsmath} 
\usepackage{amssymb} 
\usepackage{amsfonts} 
\usepackage{bm}
\usepackage{booktabs}
\usepackage{multirow}
\usepackage{caption}
\usepackage{graphicx} 
\usepackage{subfigure}
\graphicspath{{./images/}}
\usepackage{amsthm}

\usepackage{hyperref}
\hypersetup{colorlinks=true, citecolor=cyan, linkcolor=cyan}
\usepackage{xcolor} 
\usepackage[style=apa, natbib=true, giveninits=true, backend=biber, doi=false]{biblatex}
\AtEveryBibitem{\clearfield{note}}
\usepackage{appendix}

\newcounter{lccomm}

\providecommand{\keywords}[1]
{\small	
\textbf{\textit{Keywords---}} #1}
\title{Abstract, keywords and references template}
\author{Author Surname$^{1}$, Someone Else$^{2}$  \\
        \small $^{1}$University A \\
        \small $^{2}$University B \\}      
        
\def\v{\boldsymbol}
  
\newcounter{mntcomm}

\begin{document}
\newgeometry{top=0.75in, bottom=1in, left=1in, right=1in}
\title{\textbf{Global Neural Networks and The Data Scaling Effect in Financial
Time Series Forecasting}}
\author{Chen Liu\thanks{\textit{Discipline of Business Analytics, the University of Sydney Business School, Australia}}
\and Minh-Ngoc Tran\footnotemark[1]
\and Chao Wang\footnotemark[1]
\and Richard Gerlach \footnotemark[1]
\and Robert Kohn\thanks{\textit{School of Economics, UNSW Business School, Australia}}
}
\date{}
\maketitle
\vspace{-2em}
\begin{abstract}

Neural networks have revolutionized many empirical fields, yet their application to financial time series forecasting remains controversial. In this study, we demonstrate that the conventional practice of estimating models locally in data‐scarce environments may underlie the mixed empirical performance observed in prior work. By focusing on volatility forecasting, we employ a dataset comprising over 10,000 global stocks and implement a global estimation strategy that pools information across cross‐sections. Our econometric analysis reveals that forecasting accuracy improves markedly as the training dataset becomes larger and more heterogeneous. Notably, even with as little as 12 months of data, globally trained networks deliver robust predictions for individual stocks and portfolios that are even not in the training dataset. Furthermore, our interpretation of the model dynamics shows that these networks not only capture key stylized facts of volatility but also exhibit resilience to outliers and rapid adaptation to market regime changes. These findings underscore the importance of leveraging extensive and diverse datasets in financial forecasting and advocate for a shift from traditional local training approaches to integrated global estimation methods.

\end{abstract}
\restoregeometry

\newpage
\section{Introduction}\label{sec:introduction}
Volatility modeling is a central focus in financial econometrics due to its significant implications for risk management, portfolio allocation, and option pricing; see, e.g., \citet{lehar_garch_2002, brooks_volatility_2003, symitsi_covariance_2018, alves_forecasting_2024}. The Generalized Autoregressive Conditional Heteroskedasticity (GARCH) model by \citet{bollerslev_generalized_1986} and the Heterogeneous Autoregressive (HAR) model by \citet{corsi_simple_2009}, along with their extensions, have been widely adopted due to their effectiveness in capturing the volatility dynamic in financial returns. These models share a common feature of having a parsimonious structure, making it easy to perform model estimation and inference.

These econometrics models are typically fitted to individual financial time series. We refer to this approach as local training, and a model trained in this manner as a {\it local model}. While this approach allows for tailoring to the specific characteristics of each series, it fails to exploit the potential efficiency gains from utilizing cross-sectional information across multiple related series \citep{makridakis_m4_2018, montero-manso_principles_2021}. An alternative is the global training approach, which trains a single model on a panel of related time series. We refer to a model trained in this way as a {\it global model}. Recent research has increasingly highlighted the advantages of global models, which leverage shared information and cross-sectional dependence. For example, \citet{garza_timegpt-1_2023} show that a large global model trained on diverse time series datasets including retail sales, electricity consumption, transport and banking, outperforms local models by capturing common temporal patterns. \citet{smyl_hybrid_2020}, the winner of the M4 forecasting competition, uses a local-global hybrid model in which exponential smoothing is applied locally to each individual time series followed by a large neural networks trained globally across all time series. \citet{montero-manso_principles_2021} demonstrate that global forecasting methods outperform traditional local models by leveraging scalability, reduced complexity and improved generalization, especially in large and diverse datasets.

This paper explores the global approach for volatility modelling and forecasting. The first aim of the paper is to contrast econometric models to large machine learning models, examining their effectiveness as local and global frameworks for capturing and predicting market volatility.

The effectiveness of the global approach largely depends on the degree of heterogeneity between the pooled time series, where time series with low heterogeneity are more suitable for a global approach \citep{bandara_forecasting_2020,montero-manso_principles_2021}. In the context of financial markets, time series often exhibit common characteristics such as volatility clustering and mean reversion, which are prevalent across different assets and markets \citep{engle_autoregressive_1982, bollerslev_generalized_1986, diebold_measuring_2009, adrian_covar_2011}. These shared features make the global approach particularly suitable for financial time series forecasting. 

There have been several attempts in this area using econometric models to pool information across multiple time series. For example,
\citet{barigozzi_euro_2014} found that the euro area countries’ responses to monetary policy shared common components that could be effectively captured using a factor model to pool information across many time series; their method can be viewed as a global approach.
Similarly, through panel-based estimation method, \citet{bollerslev_risk_2018} demonstrated the importance of accounting for volatility spillovers across markets, highlighting the relevance of global models for managing cross-market risks. \citet{brownlees_backtesting_2021} reinforced the utility of global models in their backtesting of global growth-at-risk, while \citet{kleen_forest_2022} showed that pooling risk forecasts from multiple financial instruments provides a broader view of market volatility and improves forecasting accuracy. While these studies do not explicitly discuss the global training approach, their methods effectively pool information from multiple time series, showcasing the benefits of global-like models.

The paper's second aim is to investigate whether homogeneity among financial time series guarantees the effectiveness of the global approach for modeling and forecasting volatility.

Our third objective is to study how data size and diversity affect global models. Recent studies on scaling laws in natural language processing and computer vision show that increasing dataset size can substantially enhance predictive accuracy \citep{kaplan_scaling_2020, zhai_scaling_2022}, emphasize the importance of prioritizing the quality and scale of training data over focusing solely on model architecture. In modern finance, the availability of financial datasets has increased dramatically, creating opportunities for data-centric methods that leverage the power of NNs to uncover patterns directly from data. Despite this, the use of such methods remains limited in mainstream econometrics. Understanding how data size and diversity influence the performance of global NNs is crucial for advancing their practical adoption in finance industry.

Through extensive empirical analysis on a large-scale stock dataset, we identify four key findings:

First, less parameterized models, such as GARCH, perform poorly in global settings, with accuracy deteriorating as the number of pooled series increases. Conversely, highly parameterized models like NNs improve in accuracy as the pool size grows, a finding consistent across different NN architectures. Moreover, NNs trained on diversified datasets consistently outperform those trained on homogeneous groups, highlighting the importance of data diversity alongside size.

Second, we address the mixed views within the econometrics community on the effectiveness of NNs compared to econometric counterparts \citep{makridakis_statistical_2018}. While locally trained NNs underperform econometric models, globally trained NNs consistently outperform both their local counterparts and econometric baselines. This underscores the importance of adopting the global approach as the standard when applying NNs to financial time series forecasting.

Third, global NNs offer many practical advantages over their local counterparts, including superior forecasting accuracy, effective generalization to unseen stocks and portfolios, and reduced data and computational requirements. Once trained, they require as little as twelve months of data to accurately forecast stocks or portfolios.

Finally, global NNs demonstrate the ability to learn key stylized facts of volatility, such as clustering and leverage effects, directly from data. They exhibit robustness to outliers and adapt effectively to rapidly changing market conditions, attributes facilitated by their flexible structures and diverse training datasets.

The paper is organized as follows. Section 2 introduces the global training approach and NN architectures applied in the paper. Section 3 examines the data scaling effects of global models. Section 4 compares global NNs with local NNs and econometric baselines, focusing on VaR and ES forecasts and providing an in-depth interpretation of global NNs. The appendix includes further technical details, training configurations, and a glossary to assist readers unfamiliar with machine learning terminologies.

\subsection{Ready-to-go Python Notebook}
We provide a Jupyter notebook with a pre-trained global volatility model. Users can simply follow step-by-step instructions to forecast stock volatility, examine leverage effects, analyze responses to outliers, and compare results with stock-specific GARCH models. The notebook is available at \url{https://github.com/cqlc94/DeepVol}.

\section{Methodology}\label{sec:Methodology}
\subsection{Training schemes}
\paragraph{Local training.} Let $\v y=\{y_t : t=1, \ldots, T\}$ represent a daily time series of demeaned returns. In volatility modeling, the key quantity of interest is the conditional return variance,  $\sigma_t^2 = \operatorname{var}(y_t \mid \mathcal{F}_{t-1})$, where $\mathcal{F}_{t-1}$ contains a set of available information up to time $t-1$, which, in our case, is the past returns. A local volatility model, assuming Gaussian errors, is specified as
\begin{subequations}\label{eq:local_model}
\begin{gather}
    y_t = \sigma_t \epsilon_t, \quad \epsilon_t \overset{\mathrm{i.i.d}}{\sim} \mathcal{N}(0,1),\\
    \sigma_t = f\left(y_{1:t-1}\right), \quad t=1, 2, \ldots, T_{in}, \\
    \ell(\v y \mid \v \theta) = -\frac{1}{T_{in}} \sum_{t=1}^{T_{in}} \log\left(p\left(y_t \mid \sigma_t\right)\right).
\end{gather}
\end{subequations}
Here, $T_{in}$ represents the in-sample data size. The function $f(\cdot)$ forecasts the conditional volatility $\sigma_t$, and it could be an econometric model (e.g., GARCH) or a NN model. The negative log-likelihood $\ell(\v y \mid \v \theta)$ is the objective to minimize, where $\v \theta$ denotes all the model parameters, and $p(\cdot \mid \sigma_t)$ is a Gaussian density function with mean zero and variance $\sigma_t^2$.

Local training means estimating the model parameters using a single time series $\v y$. The resulting model is called a {\it local model}, as it is trained independently for each asset. For example, if $f(\cdot)$ is based on a GARCH(1,1) model, i.e. $f\left(y_{1:t-1}\right)=\left(\omega+\alpha y_{t-1}^2+\beta\sigma_{t-1}^2\right)^{1/2}$, then the parameters $\v \theta = (\omega, \alpha, \beta)$ are optimized specifically for each series $\v y$. 


\paragraph{Global training.} Global training uses several time series to estimate a shared set of parameters, forming a global model. We write this approach as
\begin{subequations}\label{eq:global}
\begin{gather}
    y_t^n = \sigma_t^n \epsilon_t^n, \quad \epsilon_t^n \overset{\mathrm{i.i.d}}{\sim} \mathcal{N}(0,1), \quad t = 1, 2, \ldots, T_{in}, \quad n = 1, 2, \ldots, N, \label{eq:DeepVol1} \\
    \sigma_t^n = f^*(y_{1:t-1}^n), \label{eq:DeepVol2} \\
    \ell(\v Y \mid \v \theta^*) = -\frac{1}{N T_{in}} \sum_{n=1}^{N} \sum_{t=1}^{T_{in}} \log\left(p(y_t^n \mid \sigma_t^n)\right). \label{eq:DeepVol3}
\end{gather}
\end{subequations}
Here, a single model $f^*(\cdot)$ is trained across all series, optimizing parameters $\v \theta^*$ collectively. For example, using global training to fit a GARCH(1,1) model on 10,000 stock series results in a single parameter $\v \theta^* = (\omega^*, \alpha^*, \beta^*)$, optimized over the entire 10,000 time series dataset.

In econometrics, panel data models \citep[see, e.g.,][]{chamberlain1982multivariate} share similarities with global models in that both pool information across series. Panel data models achieve this by incorporating fixed effects to account for shared characteristics across series. Typically, panel data models are part of a multivariate regression framework designed to capture interdependence among multiple series. In contrast, global models assume that the time series share common patterns but remain independent. While global training utilizes several time series during training, it produces a univariate model that forecasts each series independently. At the stage of inference, it uses only the return series of a single stock to make predictions.


\subsection{Neural Networks}
Our paper focuses on employing neural networks to explore global training, and examine how data size affect the performance of global models in financial time series forecasting. NNs are well-suited for this task due to their capacity to model complex, nonlinear relationships in data \citep{LeCun:2015pmr, zhou_informer_2021,wu_autoformer_2021}. We briefly describe recurrent neural networks (RNNs) which are designed for sequential data. The interested reader is referred to \citet{goodfellow2016deep,benidis_deep_2022} for a comprehensive introduction. 

\begin{figure}[h!]
    \centering
    \includegraphics[width=0.4\textwidth]{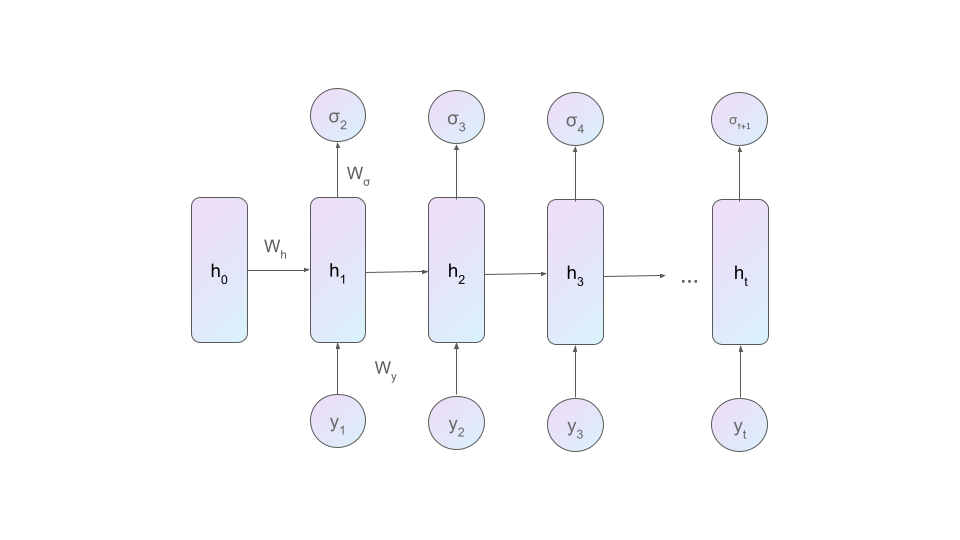}
    \caption{Graphical representation of the basic RNN.}
    \label{fig:rnn}
\end{figure}

Unlike econometric models, such as GARCH, that specify temporal relationships through simple linear functions, RNNs capture temporal dependence by leveraging latent variables as a form of memory. These latent states are recursively updated with both past memory and current input values. 
The basic RNN structure is given by
\begin{subequations}\label{eq:rnn}
\begin{gather}
{h_t} = \psi\left(W_{y} y_t + {W_{h}} {h_{t-1}} + {b_h}\right) \\
\sigma_{t+1} = \operatorname{ReLU}\left({W_{\sigma}} {h_t} + b_{\sigma}\right) + 1e^{-8},
\end{gather}
\end{subequations}
where $y_t$ is the input (the stock's return) at time step $t$, ${h_t}$ denotes the hidden state vector at time $t$, and $\sigma_{t+1}$ is the predicted volatility at time $t+1$. The weight matrices $W$ and bias vectors $b$ are the model parameters. The function $\psi$ is a non-linear activation function, typically chosen as the sigmoid function to model non-linear dependence and $\operatorname{ReLU}(x)=\max\{x,0\}$ is used in the output layer to guarantee non-negative volatility predictions. The small number $1e^{-8}$ is added for a numerical stability. Figure \ref{fig:rnn} illustrates this RNN. 

A well-documented limitation of this basic RNN model is its difficulty in effectively learning long-term dependence. During training, the gradient of the loss function with respect to the model parameters either vanishes or explodes as it propagates through time steps \citep{goodfellow2016deep}. One of the most widely-used  RNN variants designed to address this issue is the Long Short-Term Memory (LSTM) network of \citet{hochreiter_long_1997}. LSTM uses an additional hidden state vector, called a memory cell, to avoid the vanishing-exploding gradient problem. Furthermore, it introduces a gate structure to regulate the memory flow. By selectively retaining or discarding past memory at each time step, this model is particularly well-suited for datasets with complex long-range dependence and non-stationary characteristics, such as those frequently encountered in financial time series. 

The LSTM model is written as follows
\begin{subequations}
    \begin{gather}
    f_t=\psi\left(W_{f y} y_t+W_{f h} h_{t-1}+b_f\right) \\
    i_t=\psi\left(W_{i y} y_t+W_{i h} h_{t-1}+b_i\right) \\
    o_t=\psi\left(W_{o y} y_t+W_{o h} h_{t-1}+b_o\right) \\
    \tilde{C}_t=\tanh \left(W_{C y} y_t+W_{C h} h_{t-1}+b_C\right) \\
    C_t=f_t \odot C_{t-1}+i_t \odot \tilde{C}_t \\
    h_t=o_t \odot \tanh \left(C_t\right) \\
    \sigma_{t+1}=\operatorname{ReLU}\left(W_{\sigma h} h_t+b_\sigma\right)+1e^{-8}.
    \end{gather}
\end{subequations}
Here, the vectors $i_t$, $f_t$ and and $o_t$ are the input gate, the forget gate and the output gate $o_t$, respectively.
The additive structure of the memory cell state $C_t$ helps mitigate the vanishing and exploding gradient issue. The matrices $W$ and vectors $b$ are the model parameters. 
We also consider another RNN variant, the Gated Recurrent Unit (GRU) of \citet{chung_empirical_2014} described in the Appendix \ref{sec:training_configuration}.

\subsection{Dataset} 
We collected the daily closing prices of more than 12000 stocks from 10 exchanges through the Reuters Refinitiv Workspace, covering the decade from January 1, 2014, to January 1, 2024. The time series for each stock was split into training, validation and testing periods following a 60\%/20\%/20\% split. The training period spans from 2014 to 2019, the validation period spans from 2020 to 2021, and the testing period spans from 2022 to 2023. To ensure the local models can be effectively trained, stocks with fewer than 1,260 (five years) observations during the training period are excluded, leaving a total of 11,771 stocks. Data summary statistics are in Appendix \ref{sec:Data}. 

\subsection{Experiment setup for data scaling effect study}
This section describes the study of the effect of data size on the performance of global models, i.e., how out-of-sample predictive performance changes as the number of pooled stock series varies. Specifically, we train multiple global NNs while holding all other model hyperparameters constant, varying only the number of stock series in the training set from 10 to 10,240, doubling the data size at each step, and comparing their predictive accuracy.

We assess the models' predictive accuracy in two forecasting scenarios: \textit{supervised forecasting} and \textit{zero-shot forecasting}. In supervised forecasting, models predict the out-of-sample period for stock series whose in-sample period is included in the training process. In zero-shot forecasting, models predict the out-of-sample period for unseen stock series whose in-sample data are excluded from training. Of the 11,771 stock series in our dataset, 10,240 are allocated for training and evaluating supervised forecasting (referred to as \textit{training series}), while the remaining 1,531 are reserved for evaluating zero-shot forecasting (referred to as \textit{unseen series}).

Zero-shot forecasting performance is evaluated on the 1,531 unseen series for all models. For supervised forecasting, evaluation is restricted to the 10 stock series in the smallest training set rather than all 10,240 training series. This is because, when studying the data scaling effect, the first global model is trained on only 10 stocks; evaluating this model across all 10,240 training series would essentially assess its zero-shot capability rather than its supervised performance, which is not the intended focus of this evaluation.

It is important to note that when a stock series is used for training, only its in-sample period is included. Unless otherwise stated, all results in this paper are based on one-day-ahead forecasts during the out-of-sample testing period.

As this study involves training a large number of NNs, we standardized the hyperparameters and training configurations to ensure fair comparisons. Details of the training configurations are provided in Appendix \ref{sec:training_configuration}.

\subsection{Evaluation metrics}
This section introduces the evaluation metrics employed in our study. A comprehensive evaluation framework is essential to assess the performance of volatility models from both statistical and economic perspectives. To achieve this, we employ the metrics that measure the accuracy of volatility forecasts, the effectiveness in risk management applications, and the statistical significance of model performance at the individual stock level. By combining these metrics, we aim to provide a robust assessment of the models’ predictive abilities, economic utility, and practical relevance in financial applications.

\paragraph{Statistical metrics:}
To assess the fitness of a volatility model to a return series, we employ the Negative Log-Likelihood (NLL), also called the Partial Predictive Score in the statistics literature,
a widely used metric in statistical modeling for evaluating predictive performance. NLL quantifies the negative log-likelihood of observing the return series given the model’s volatility forecast, with smaller values indicating better performance. The NLL is defined as:
\begin{equation}
\text{NLL}:=-\frac{1}{T_{\text {test }}} \sum_{y_t \in D_{test}} p\left(y_{t} \mid y_{1: t-1},\widehat{\theta}\right),
\end{equation}
where $\widehat{\theta}$ are the estimated model parameters.

\paragraph{Economic utility:}
In addition to statistical fitness, we assess the economic utility of volatility models in risk management, a crucial application of volatility modeling. Specifically, we focus on forecasting VaR and ES, two key risk measures widely used in financial regulation and recommended by the Basel Accord. The $\alpha$-level VaR represents the $\alpha$ quantile of the return distribution, while the $\alpha$-level ES corresponds to the conditional expectation of returns exceeding the corresponding VaR.

To evaluate the accuracy of VaR forecasts, we use the quantile loss function \parencite{koenker_regression_1978}
\begin{equation}
\text{Qloss}_\alpha := \frac{1}{T_{\text{test}}}\sum_{y_t \in D_{test}}\left(\alpha-I\left(y_t<Q_t^\alpha\right)\right)\left(y_t-Q_t^\alpha\right),
\end{equation}
where $Q_t^\alpha$ is the forecast $\alpha$-level VaR at time $t$. The quantile loss function is strictly consistent \parencite{fissler_higher_2016}, meaning the expected loss is minimized when the forecast accurately predicts the true quantile. The model with the lowest quantile loss is therefore preferred for VaR forecasting.

While no strictly consistent loss function exists for ES in isolation, \citet{fissler_higher_2016} demonstrate that ES and VaR are jointly elicitable, allowing for their joint evaluation using specific loss functions. One such function, based on the Asymmetric Laplace (AL) distribution, is strictly consistent for jointly assessing VaR and ES \parencite{taylor_forecasting_2019}. The AL-based joint loss function is defined as:
\begin{equation}
\text{JointLoss}_\alpha := \frac{1}{T_{\text{test}}}\sum_{y_t \in D_{test}}\left(-\log \left(\frac{\alpha-1}{\mathrm{ES}_t^\alpha}\right)-\frac{\left(y_t-Q_t^\alpha\right)\left(\alpha-I\left(y_t \leq Q_t^\alpha\right)\right)}{\alpha \mathrm{ES}_t^\alpha}\right),
\end{equation}
where $\mathrm{ES}_t^\alpha$ is the forecast $\alpha$-level ES at time $t$. Consistent with econometric practice in risk management, we report both quantile loss and joint loss at the 1\% and 2.5\% levels.

\paragraph{Statistical significance:}
Beyond assessing aggregate performance across stocks, we evaluate the statistical significance of global model performance at the individual stock level. To achieve this, we use the Model Confidence Set (MCS) procedure introduced by \citet{hansen_model_2011}.
The MCS identifies a Superior Set of Models (SSM), defined as the subset of models that demonstrate equal predictive accuracy based on sequential hypothesis testing.

Let $\mathcal{M}$ denote the set of competing models. For models $i$ and $j$ in $\mathcal{M}$, the relative loss is defined as $d_{i,j,t} = L_{i,t} - L_{j,t}$, where $L_{i,t}$ represents the performance loss of model $i$ at time $t$. The MCS tests the null hypothesis:
\begin{equation}
H_{0}: \mu_{i,j}=0, \quad \text{for all } i, j \in \mathcal{M},
\end{equation}
where $\mu_{i,j} = \mathbb{E}(d_{i,j,t})$ is the expected relative loss. A model is excluded from the SSM if the null hypothesis of equal predictive accuracy is rejected. The remaining models in the SSM are those for which $H_0$ cannot be rejected. The MCS assigns a $p$-value $p_i$ to each model $i \in \mathcal{M}$, with higher $p$-values indicating a higher likelihood of inclusion in the SSM.
The interested reader is referred to \citet{hansen_model_2011} for more details.

\section{Data Scaling Effect}\label{sec:scaling_law}
This section investigates data scaling effect of global models. We use test NLL as the primary predictive score in this section; other predictive scores more relevant to risk management are examined in Section \ref{sec:Global Volatility Model}.

\subsection{Local GARCH versus Local NNs}
We first study the performance of NNs and the econometrics models when the local training approach is used. Our aim is to address the ongoing debate within the econometrics community regarding the effectiveness of NN models in financial time series forecasting \parencite{makridakis_statistical_2018}. For NN models, we choose basic RNN, LSTM and GRU. Inspired by \citet{hansen_forecast_2005} who decisively confirm the performance of GARCH(1,1), we choose it to represent the econometrics models; other econometric volatility models are considered in Section \ref{sec:Global Volatility Model}.

Table \ref{tab:localgarch_vs_localnn} reports the average out-of-sample NLL for local NNs, along with their respective win rates against the GARCH(1,1) model across 11,771 stocks. Additional results for risk management metrics and other econometric baseline models are presented in Table \ref{tab:risk_management_stock} and \ref{tab:risk_management_stock_mcs}. The results indicate that even the best-performing NN architecture, the LSTM, outperforms the GARCH(1,1) model in only 34\% of the stocks and on average underperform compared to the GARCH(1,1) model. This confirms the finding in \citet{makridakis_statistical_2018} that ML models often do not outperform simple statistical models in financial time series forecasting. 

\begin{table}[h!]
\centering
\caption{Performance comparison of local models: This table presents a performance comparison between various NN models and the GARCH(1,1) baseline across 11,771 individual stock series in the local training setting. The NLL column shows the out-of-sample average NLL across all stocks. The Win Rate column shows the percentage of stocks for which a given NN model outperforms the GARCH(1,1) baseline.}
\begin{tabular}{rcc}
\toprule
      & NLL   & Win Rate \\ \midrule
GARCH & 2.261 & -        \\
RNN   & 2.273 & 17\%     \\
GRU   & 2.269 & 29\%     \\
LSTM  & 2.266 & 32\%     \\ \bottomrule
\end{tabular}
\label{tab:localgarch_vs_localnn}
\end{table}
These results stand in stark contrast to the widespread success of NNs in other fields. Here, we attempt to explain this discrepancy and, in subsequent sections, propose to correct this perception through the use of global training. Financial time series are traditionally treated as heterogeneous, leading to the use of local training approaches where models are fitted individually for each series. As a result, this approach only makes use of limited data— typically around several thousand observations per stock — which is insufficient for effectively training NNs, as they generally require large datasets to learn generalized patterns. In contrast, econometric models excel in financial time series forecasting even with limited data because they incorporate theoretical patterns into their structure. For instance, the GARCH model is designed to capture the clustering effect, while GJR \citep{glosten_relation_1993} and EGARCH \citep{nelson_conditional_1991} models are additionally designed to capture the leverage effect. These structural assumptions make econometric models particularly well-suited for data-scarce environments, offering a level of robustness that is challenging for black-box neural networks to achieve under similar conditions.

\subsection{Global GARCH}
\begin{figure}[h!]
    \centering
    \subfigure[Supervised]{\includegraphics[width=0.48\textwidth]{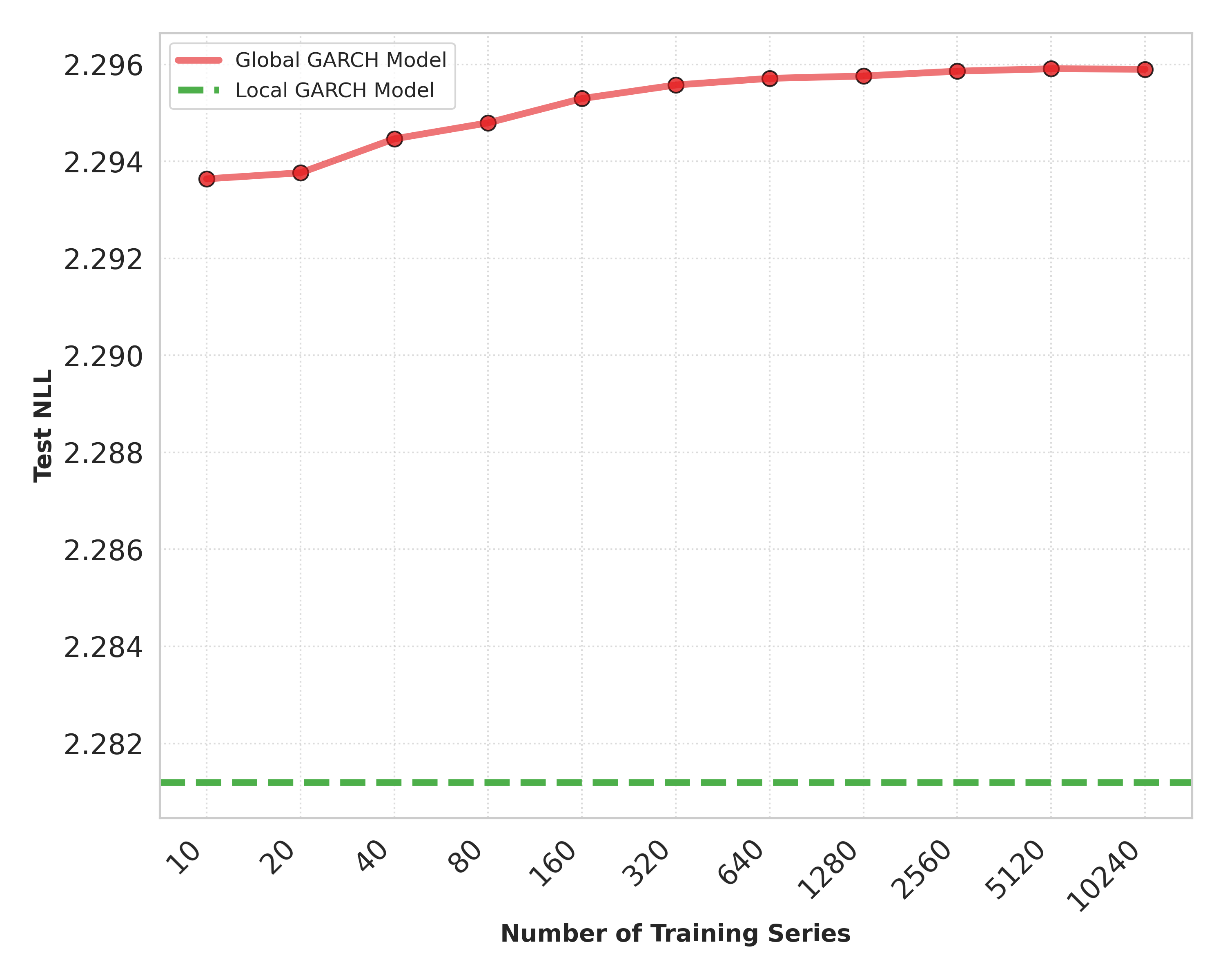}
    \label{fig:garch_data_scaling_supervised}}
    \subfigure[Zero-shot]{\includegraphics[width=0.48\textwidth]{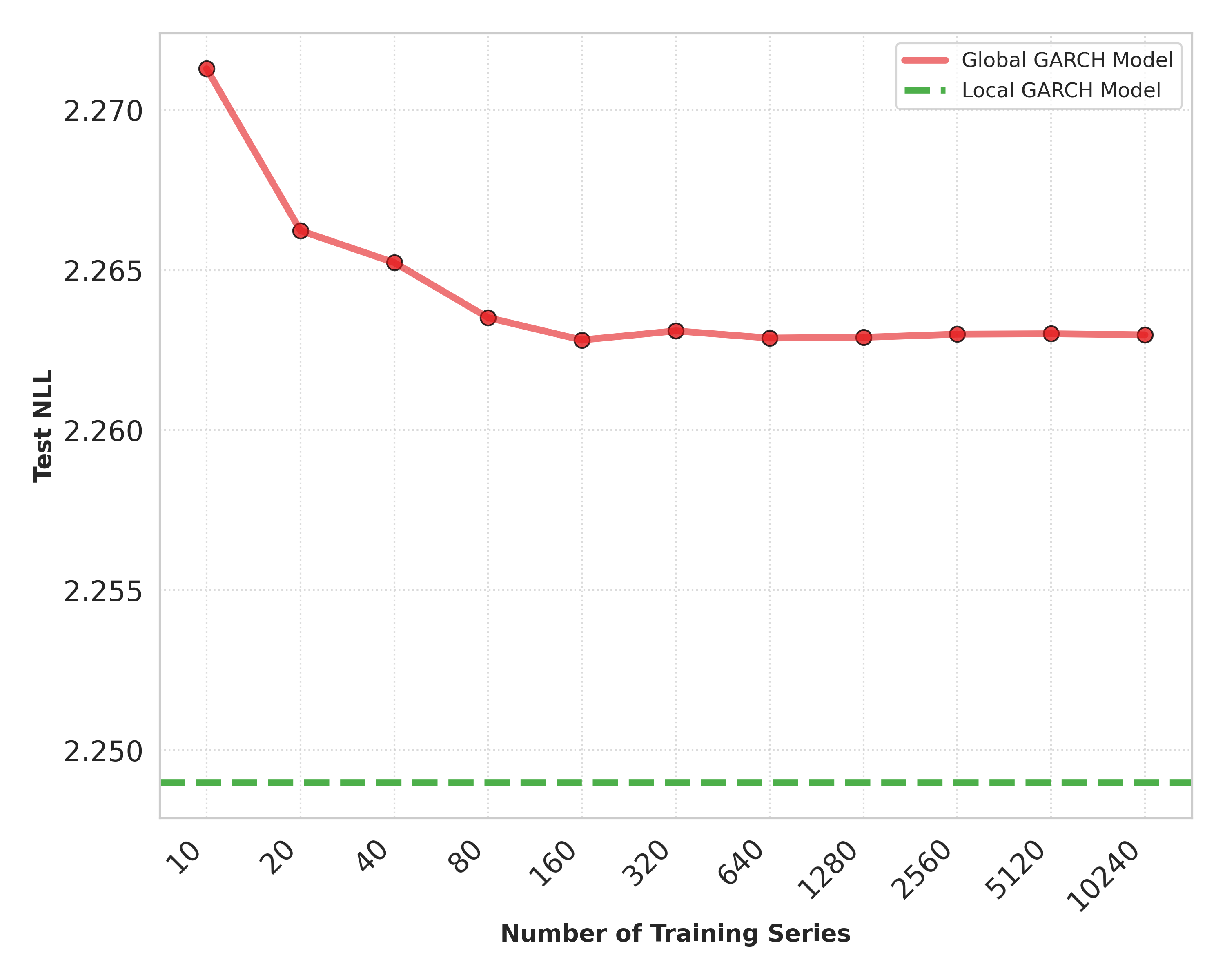}   
    \label{fig:garch_data_scaling_zeroshot}}
    \caption{GARCH data scaling effect: For supervised forecasts, the models are evaluated on 10 training stocks. For zero-shot forecasts, the models are evaluated on 1531 unseen stocks.}
    \label{fig:garch_data_scaling}
\end{figure}

However, when econometric models are applied as global models, their performance characteristics change considerably. Figure~\ref{fig:garch_data_scaling} plots the data-size scaling effect of the global GARCH model and the local baseline. Even when the global GARCH model is trained on only 10 stock series, it performs substantially worse than its local counterparts. This decline in performance arises from the global GARCH model's inability to capture the increased heterogeneity across different stocks.

The GARCH model is effective for individual-level stock series because it is tailored to identify stock-specific patterns. However, when applied globally, the assumption that a single set of parameters can describe all stocks breaks down, as the volatility behaviors across stocks differ significantly that 
the GARCH's parsimonious structure cannot adequately account for such heterogeneity. Financial time series often have distinct characteristics shaped by factors such as industry sector, market conditions, and stock-specific dynamics. As a result, the simplicity of the GARCH model, which is suitable for individual stocks, is unsuitable as a global model, as it cannot simultaneously capture the diverse volatility patterns of multiple assets. 
One potential approach to address this limitation could involve coupling a random-effects framework with GARCH, where the common fixed-effect structure is retained across stocks, but random-effect parameters allow for stock-specific variations. However, we do not pursue this idea further in this paper.

As the number of stock series used for training increases, the global GARCH model shifts from capturing stock-specific idiosyncrasies to representing the aggregate characteristics of the pooled dataset. In supervised settings, this transition leads to a decline in forecast accuracy for the 10 training stocks, as the model’s ability to reflect unique stock-level features diminishes. Conversely, in zero-shot settings—where the model forecasts stock series not included in the training set—the global model achieves improving performance by leveraging the aggregate characteristics of the broader dataset.

Table \ref{tab:garch_parameters} in Appendix \ref{sec:garch_parameters} presents the estimated parameters for these 11 global GARCH models. The parameter estimates stabilize once the model is trained on more than 160 stock series, marking a clear transition from modeling stock-specific features to capturing general patterns shared across the dataset. Beyond this point, the global GARCH model predominantly reflects the aggregate dynamics of all the stock series rather than the unique characteristics of individual or group of stocks.

In conclusion, while the structured restrictions imposed by econometric models are beneficial in local training settings, these restrictions and the lack of model complexity limit their performance in global settings. This result underscores the limitations of traditional econometric models in generalizing across diverse time series and highlights the need for more complex models capable of effectively handling the heterogeneity present in financial data. The following section examines the performance of global NNs.

\subsection{Global NNs}
\subsubsection{Data scaling effect}
Unlike econometric models, NNs perform particularly well in data-rich environments due to their flexible structure and expressive power, which allows them to capture complex relationships in large and diverse datasets. In this section we demonstrate the performance of global NNs and examine their data scaling effect.

To assess the sensitivity of data scaling effect to NN architectures, we study three NN architectures: RNN, GRU, and LSTM. 
Given the structural difference between these networks, simply using the same number of hidden units results in different model complexities. To ensure a valid comparison, we control for model complexity by equalizing the number of parameters across architectures \citep{kaplan_scaling_2020, zhai_scaling_2022}. This results in RNN, GRU, and LSTM models with 21, 12, and 10 hidden units respectively, each having approximately 500 parameters.

\begin{figure}[h!]
    \centering
    \subfigure[Supervised]{\includegraphics[width=0.45\textwidth]{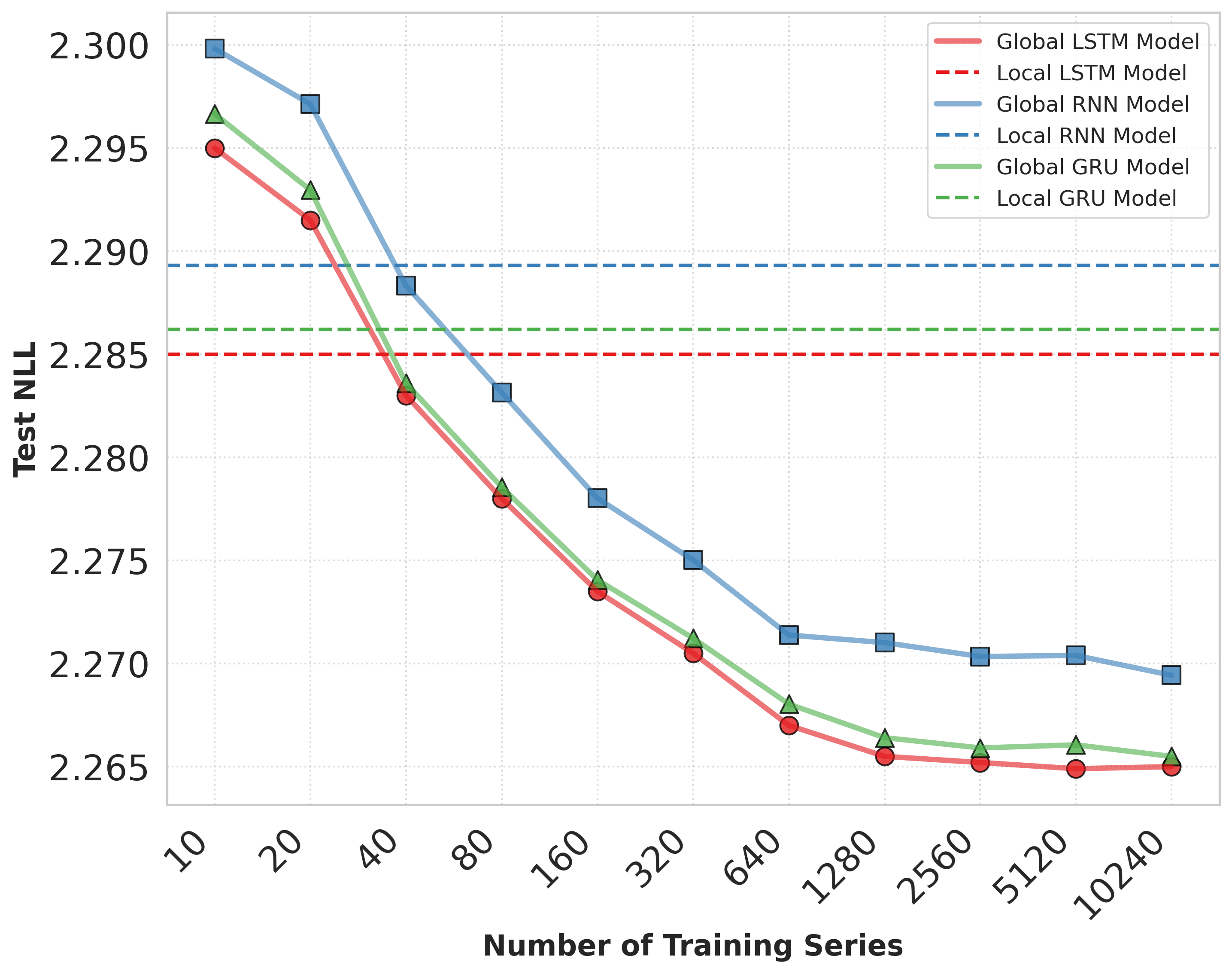}
    \label{fig:lstm_data_scaling_supervised}}
    \subfigure[Zero-shot]{\includegraphics[width=0.45\textwidth]{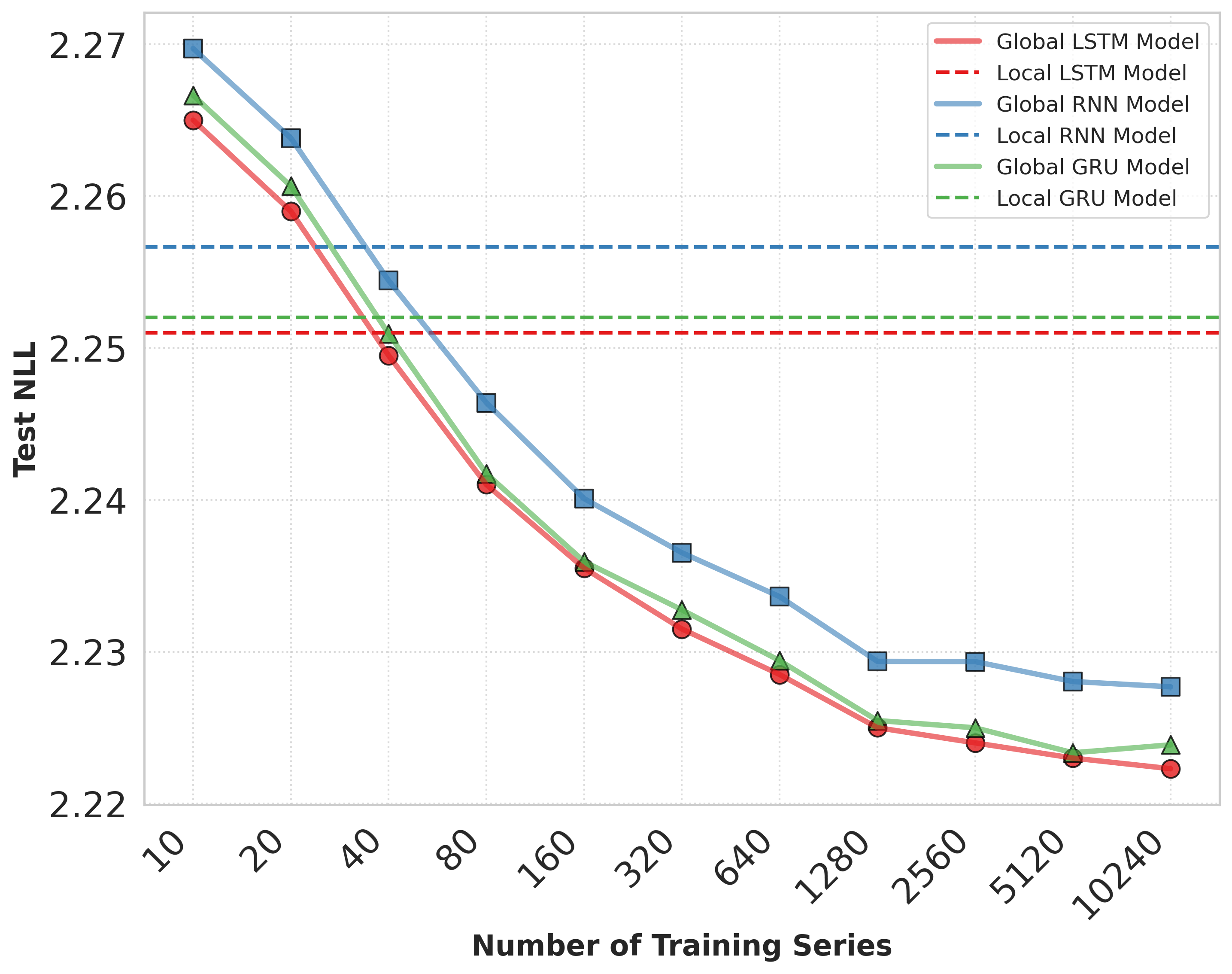}
    \label{fig:lstm_data_scaling_zeroshot}}
    \caption{NNs data scaling effect: For supervised forecasts, the models are evaluated on 10 training stocks. For zero-shot forecasts, the models are evaluated on 1531 unseen stocks.}
    \label{fig:lstm_data_scaling}
\end{figure}

Figure \ref{fig:lstm_data_scaling_supervised} shows the performance of global NNs and local NNs for both supervised and zero-shot forecasting. Unlike global econometric models, global NN models quickly outperform their local counterparts once the number of training series reaches 40. The contrasting performance of global GARCH and global NNs highlights a fundamental difference in their nature: econometric models rely on patterns imposed through prior knowledge, whereas NNs leverage their flexible structure and expressive power to uncover such patterns directly from the data. However, the flexibility of NNs comes with a requirement for substantial data to avoid overfitting noise or idiosyncratic patterns that do not generalize to out-of-sample periods. When trained on individual stock series, local NNs are particularly prone to this overfitting, which degrades their out-of-sample performance, making them inferior to simpler, well-constrained econometric models like GARCH. By pooling data from multiple stocks, global NNs overcome this limitation by learning shared temporal dynamics and capturing broader market trends that local models often miss. The increased diversity and volume of pooled data enable global models to form a more robust representation of market behavior, accommodating volatility fluctuations and other stochastic factors. Consequently, predictive accuracy improves significantly as the number of series included in training increases.

While the variation in NN architectures result in differences in forecast accuracy, the data scaling effect consistently holds. Notably, in terms of NN performance, the scale of the data plays a far more critical role than the choice of model architecture.

The results also demonstrate a critical benefit of global NNs in a zero-shot setting, where the models predict stock series not included in the training set. The global models consistently outperform local models trained directly on the target stocks. This result underscores two important insights: first, that stock series exhibit significant similarities in volatility and temporal patterns, which supports the view that they can be treated as homogeneous for modeling purposes; and second, that NNs can effectively capture these shared patterns, enabling better generalization to unseen data. 

These data scaling effects are not unique to financial applications. Similar effects are well-documented in other domains, such as natural language processing \citep{kaplan_scaling_2020, zhai_scaling_2022}, where NNs trained on large and diverse text corpora demonstrate superior performance and generalization compared to those trained on smaller, domain-specific corpora. This phenomenon is evident in the success of large language models like ChatGPT. The same principles should apply to financial time series. Consider stocks A and B. Suppose a specific pattern appears in the in-sample period of stock A and the out-of-sample period of stock B. If separate local models are trained for stocks A and B, the model for stock B would fail to recognize this pattern during out-of-sample forecasting. By contrast, a global model trained on pooled data from both stocks can learn the pattern from stock A and leverage it to improve predictions for stock B.

Despite this, stock series in finance are often modeled individually as heterogeneous, which limits the data available for NNs to identify robust and generalizable patterns. This practice is fundamentally at odds with the nature of NN models and underutilizes their strengths. We argue that, if one intends to treat stocks as heterogeneous series and train models locally, NN models may not be a suitable approach. Conversely, if NN is to be applied to financial time series, a global approach—pooling data from multiple stocks—should be considered as the standard.

\subsubsection{The effect of data diversity}
Stocks from the same country or industry sector are expected to exhibit higher levels of homogeneity, raising the question of whether grouping such stocks for modeling enhances forecast accuracy. To investigate this, the 11,771 stocks are divided into groups based on their country of origin and industry classification. Separate group-specific models are trained for each group, meaning a global model is trained using only stocks from that group. These group-specific models are then compared to global models trained on more diverse datasets. To ensure a fair comparison and eliminate the effect of data size, the global models trained on diverse datasets are matched to the size of the corresponding stock group. For example, for the China stock group containing 3,469 series, the corresponding global model is trained on 3,469 diversified stocks randomly selected from the 11,771 stocks.

Table \ref{tab:group_stocks} reports the performance comparison. The results indicate that the global model, trained on a similarly sized but more diverse dataset, generally outperforms the group-specific models. This superior performance can be attributed to the global model’s exposure to a more diverse dataset, enabling it to capture patterns shared across individual segments. These findings underscore the importance of data diversity as a solution to the data size bottleneck. Future research could be conducted by combining financial time series across different asset classes. Moreover, the results again highlight that, when employing flexible models such as neural networks, stock series should be effectively treated as homogeneous and modeled together for optimal performance.

\begin{table}[h!]
\centering
\caption{Model performance by country and industry}
\label{tab:group_stocks}
\begin{tabular}{lccc}
\toprule
 & Group Model & Global Model & NStocks \\
\midrule
\textbf{Country} \\
China (mainland) & 2.291 & \textbf{2.287} & 3469 \\
United States & 2.319 & \textbf{2.310} & 2146 \\
Japan & \textbf{1.950} & 1.951 & 1988 \\
India & 2.310 & \textbf{2.301} & 834 \\
United Kingdom & 2.157 & \textbf{2.146} & 573 \\
\midrule
\textbf{Industry} \\
Industrials & 2.181 & \textbf{2.179} & 2459 \\
Consumer Discretionary & 2.237 & \textbf{2.229} & 1633 \\
Information Technology & 2.337 & \textbf{2.321} & 1550 \\
Materials & 2.207 & \textbf{2.201} & 1396 \\
Financials & 2.215 & \textbf{2.169} & 1202 \\
\bottomrule
\end{tabular}
\end{table}

\section{Universal Volatility Model}\label{sec:Global Volatility Model}
The previous section examined the data scaling effects of NNs, and confirmed the superior performance of global NNs for financial volatility modelling. This section explores the economic utility and statistical significance of a globally-trained stock volatility model for real-world financial applications, along with a detailed interpretation of its characteristics. This global model employs an LSTM architecture with 10 hidden units and is trained on a dataset of 10,240 stock series. We refer to this trained global LSTM model as the {\it universal volatility model}.

\subsection{Data scarcity and temporal importance}\label{sec:Data scarcity and temporal importance}
Data scarcity is a persistent challenge in time series modeling, especially in business and economic applications. For example, data sets such as annual GDP series or daily stock returns often contain at most only a few thousand observations, which is considered small in successful machine learning applications. A related challenge arises with newly listed stocks which have insufficient historical data for a meaningful modeling. Traditional econometric models, such as GARCH, typically require at least two thousand observations for reliable parameter estimation \citep{nguyen_recurrent_2022}, making them less effective in such scenarios. In contrast, the ability to produce zero-shot forecasts of global models allows them to overcome data scarcity, enabling accurate volatility forecasts for newly listed stocks with fewer observations. 

To investigate the performance of universal volatility model in such data-scarce scenarios, we analyze the impact of input series length on its forecast accuracy. We adopt the remove-and-predict method, originally developed for computer vision tasks \citep{samek_evaluating_2017}, in which pixels or regions are systematically removed from input images to identify the features that contribute the most to the prediction accuracy of a trained model.
We extend this idea to assess the importance of the input length in stock series data. First, we evaluate the universal model using a rolling-window forecasting scheme with a fixed window size of 504 observations (equivalent to two trading year). That is, the model uses the past 504 return observations to make one-day-ahead volatility forecasts, and we compute the negative log-likelihood NLL for all the unseen stocks. Next, to study the effect of input length, we systematically reduce the input window size and recalculate the NLL to observe how changes in input length influence model performance. 
We define the temporal importance (TI) of size $k$ as
\begin{equation}
\text{TI}_k = 100\dfrac{\text{NLL}_{504} - \text{NLL}_{k}}{\text{NLL}_{504}},
\end{equation}
where $\text{NLL}_k$ is the NLL of the universal model evaluated with a fixed window size of $k$ on the unseen stocks.
The value of $\text{TI}_k$ indicates the relative performance of the input window of size $k$ compared to the 504-observation window. This approach allows us to quantify the contribution of various past observations to the prediction accuracy of the universal model.

\begin{figure}[h!]
    \centering
    \includegraphics[width=0.5\linewidth]{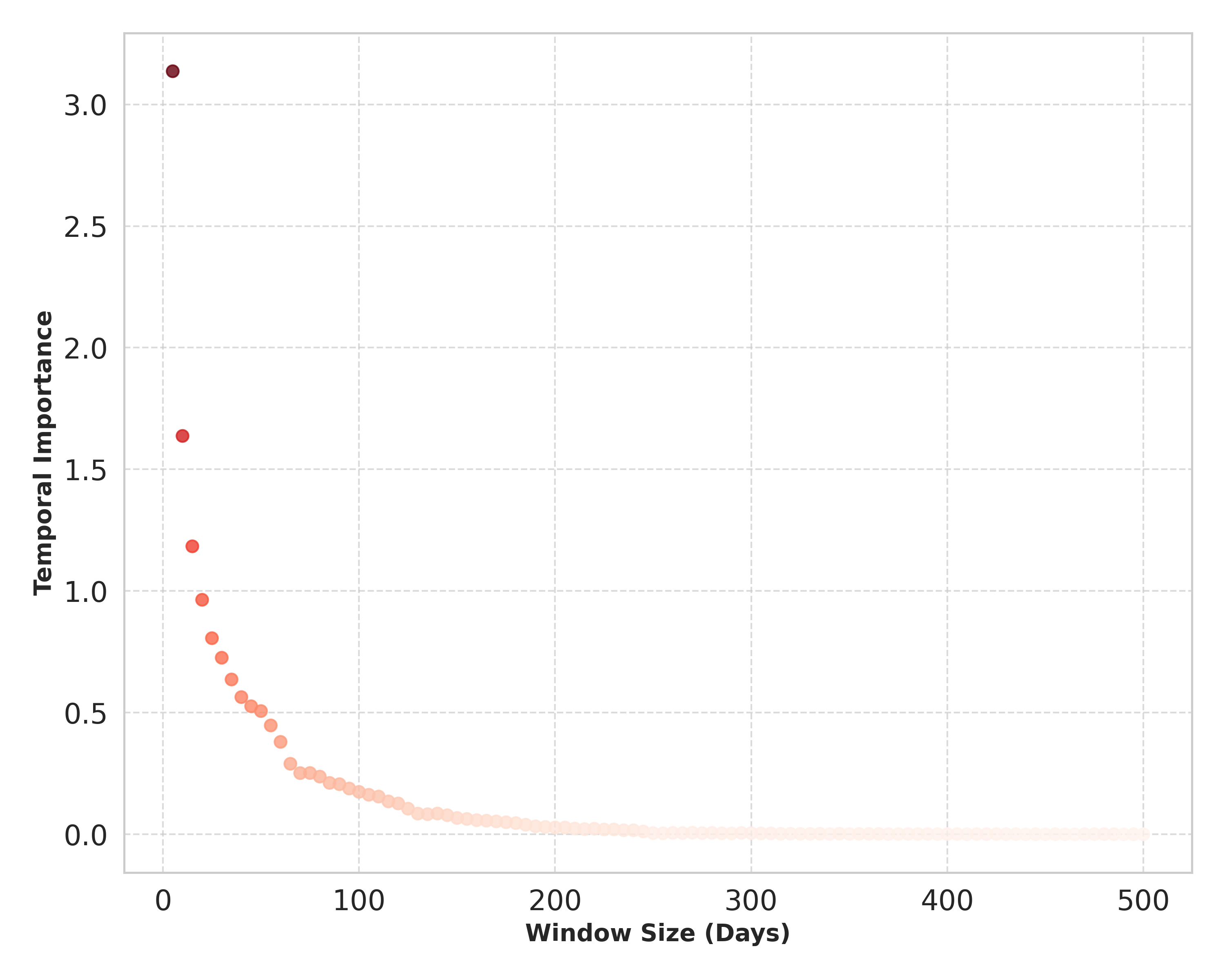}
    \caption{Temporal importance for the universal LSTM model.}
    \label{fig:time_importance}
\end{figure}

Figure \ref{fig:time_importance} plots the temporal importance of the universal LSTM model for different window sizes $k$. The results indicate that the most critical observations are typically concentrated within the past four months, with the importance of historical observations decreasing almost exponentially as they become more distant. Observations from more than twelve months prior have negligible importance and almost no impact on forecast accuracy. 
 
These findings underscore a practical advantage of universal volatility models in real-time applications for rapidly evolving financial markets. Unlike local econometric models, which generally require eight years of daily observations for reliable parameter estimation and forecasting, once trained, global NNs can directly provide accurate volatility forecasts using as little as twelve months of data, even for stocks not included in the training set. This ability is particularly valuable for newly listed stocks, where historical data is inherently limited and insufficient to fit local models effectively.

\subsection{Financial risk forecasts}
We now assess the economic utility of the universal LSTM model in risk management, using VaR and ES as the risk metrics. All local baselines—including local GARCH, local GJR, local EGARCH, and local LSTM—are evaluated using an expanding-window forecasting scheme, leveraging the full historical observations starting from 01/01/2014. In contrast, drawing on insights from the temporal importance analysis in Section \ref{sec:Data scarcity and temporal importance}, the universal LSTM is evaluated using a rolling-window forecasting scheme, which relies on only the past twelve months (252 days) of observations.  

Table \ref{tab:risk_management_stock} reports the average risk metrics for the models under consideration. Consistent with the earlier findings, the local LSTM model underperforms the econometric baselines on average. However, the Model Confidence Set results in Table \ref{tab:risk_management_stock_mcs} reveal a more nuanced picture: the local LSTM is more frequently included in the Superior Set of Models (SSM) compared to the GARCH and GJR models. This indicates that the performance of the local LSTM model is highly stock-dependent: while it can achieve strong results for certain stocks, it may perform poorly for others. In contrast, econometric models exhibit more consistent performance across stocks. The dual behavior of the local LSTM—lower average predictive accuracy but higher inclusion in the SSM when it performs well—helps explain why, despite skepticism within the econometrics community, neural network models are often reported to be superior in some studies.
The existing literature tends to focus primarily on cases where machine learning models perform well, potentially overlooking their inconsistent results across broader contexts.

\begin{table}[h!]
\centering
\begin{tabular}{lccccc}
\toprule
 & GARCH & GJR & EGARCH & LSTM & Universal LSTM \\
\midrule
NLL & 2.259 & 2.253 & 2.247 & 2.278 & \textbf{2.229} \\
QLoss 1\% & 0.102 & 0.096 & 0.094 & 0.113 & \textbf{0.082} \\
QLoss 2.5\% & 0.193 & 0.185 & 0.182 & 0.201 & \textbf{0.157} \\
JointLoss 1\% & 3.291 & 3.278 & 3.269 & 3.385 & \textbf{3.060} \\
JointLoss 2.5\% & 3.002 & 2.987 & 2.973 & 3.083 & \textbf{2.795} \\
\bottomrule
\end{tabular}
\caption{Risk metrics of the universal LSTM model and local baselines: The reported values are averages across all 11,771 stocks. For all metrics, lower values indicate better performance.}
\label{tab:risk_management_stock}
\end{table}

\begin{table}[h!]
\centering
\begin{tabular}{lccccc}
\toprule
 & GARCH & GJR & EGARCH & LSTM & Universal LSTM \\
\midrule
NLL & 2023 & 1868 & 3105 & 2773 & \textbf{8794} \\
    & (0.106) & (0.095) & (0.187) & (0.170) & \textbf{(0.685)} \\
QLoss 1\% & 2493 & 2087 & 3489 & 2621 & \textbf{8541} \\
          & (0.134) & (0.109) & (0.210) & (0.153) & \textbf{(0.661)} \\
QLoss 2.5\% & 2398 & 1989 & 3373 & 2842 & \textbf{8443} \\
            & (0.133) & (0.103) & (0.209) & (0.171) & \textbf{(0.654)} \\
JointLoss 1\% & 2534 & 2190 & 3440 & 2419 & \textbf{8636} \\
              & (0.140) & (0.113) & (0.208) & (0.139) & \textbf{(0.670)} \\
JointLoss 2.5\% & 2395 & 2046 & 3311 & 2546 & \textbf{8887} \\
                & (0.131) & (0.104) & (0.196) & (0.145) & \textbf{(0.693)} \\
\bottomrule
\end{tabular}
\caption{MCS of the universal LSTM model and local baselines: The table reports the frequency each model is included in the set of superior models 5\% significance level. The numbers in parentheses are the averaged $p$-value across all 11,771 stocks. The higher p-values indicate a higher likelihood of inclusion in the SSM.}
\label{tab:risk_management_stock_mcs}
\end{table}

The universal LSTM, even using only the past twelve months of observations, demonstrates consistent and statistically significant improvements in out-of-sample performance for risk management applications compared to the local LSTM and econometric baselines. At the individual stock level, as reported in Table \ref{tab:risk_management_stock_mcs}, the universal model also performs robustly, being included in the SSM for the majority of stocks and achieving significantly higher $p$-values. Notably, in our experiments, all econometric models are re-estimated daily during out-of-sample periods, while the global LSTM does not undergo any retraining, demonstrating its robustness to distribution shifts.

The total training time for 10,240 local LSTM models is 1382.4 minutes, compared to just 5.3 minutes for a single universal model trained on the same 10,240 series. Both models use identical hyperparameter settings and training configurations. While NNs are often criticized for high computational costs, this applies only to local models, where training time scales linearly with the number of series. In the finance industry, where a large number of assets and portfolios require modeling and frequent retraining, local NNs are computationally infeasible in practice. By contrast, a single global NN delivers accurate and consistent volatility forecasts for any stock or portfolios with just a few minutes of training or updating, offering a scalable and efficient solution.

\subsection{Model interpretation}
This section evaluates the universal LSTM model in more detail, providing additional interpretation and insights to better understand its behavior and performance.

\subsubsection{Leverage effect}
The leverage effect is a key stylized fact of stock volatility, describing the phenomenon where negative shocks to a stock’s return increase its volatility more than positive shocks of the same magnitude. This asymmetry is explicitly incorporated into the structure of GJR and EGARCH models. This section investigates whether global NNs can also capture the leverage effect.

The leverage effect is commonly measured by the news impact curve (NIC) \citep{engle_measuring_1993}, which illustrates how unexpected returns influence the volatility of an asset. For the GARCH family models such as the GJR, the NIC can be derived analytically due to their simple functional forms, enabling a direct representation of the relationship between shocks and volatility changes. In contrast, NNs lack such predefined functional forms, making it impossible to directly derive the NIC. To overcome this challenge, we adopt a simulation-based approach to approximate the NIC for NNs. Specifically, we provide the model with an input sequence of length 252 (representing the past 12 months of inputs) consisting entirely of zeros, except for the last observation, which is varied from -5 to 5. This allows us to examine how the model’s output, which is the conditional volatility, responds to these changes.

\begin{figure}[h!]
    \centering
    \includegraphics[width=0.5\linewidth]{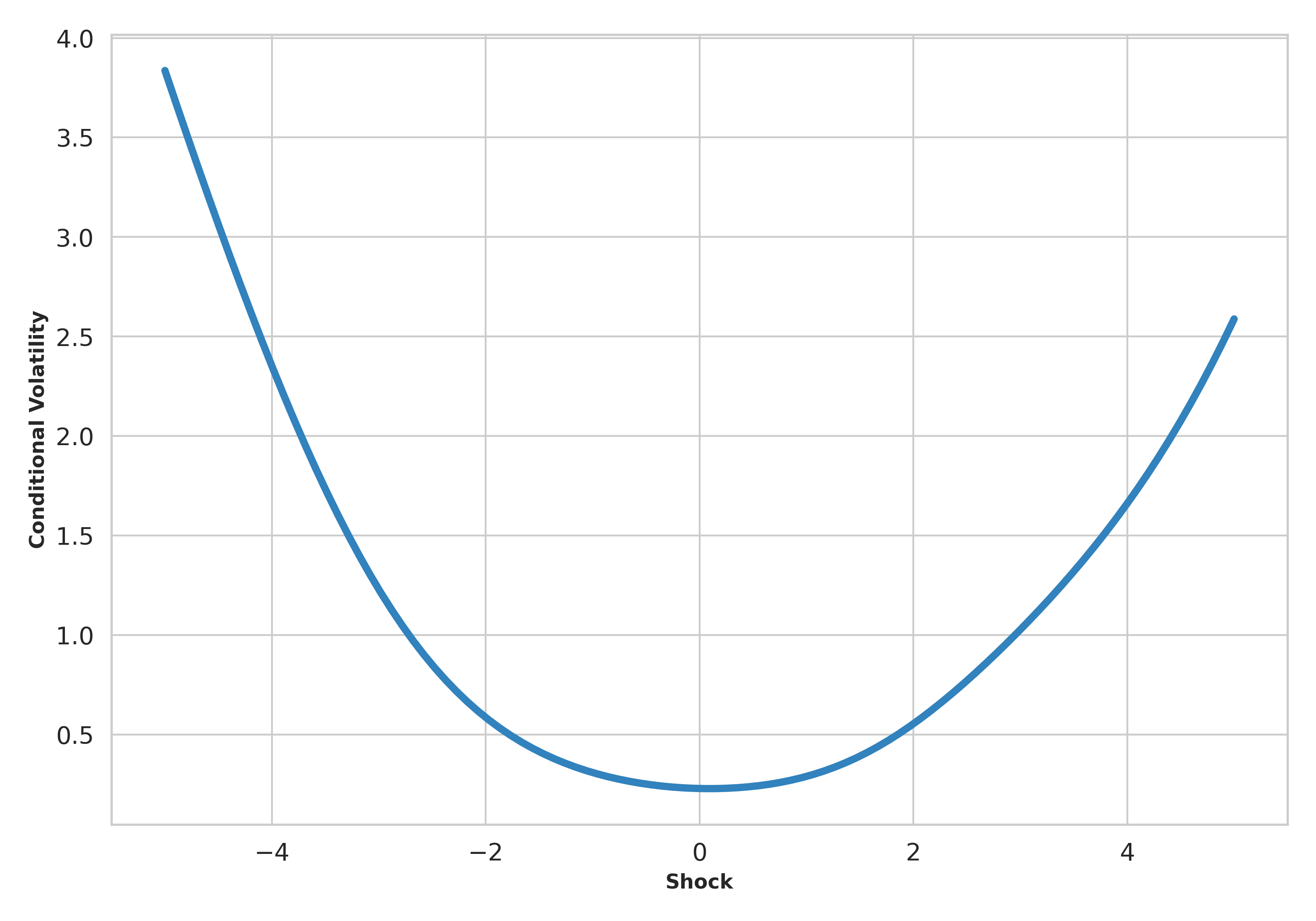}
    \caption{News impact curve of the universal LSTM model.}
    \label{fig:news_impact_curve}
\end{figure}

Figure \ref{fig:news_impact_curve} plots the NIC for the universal LSTM model. Despite lacking a predefined structure, the model effectively captures the leverage effect. It produces an asymmetric response of volatility to positive and negative shocks, with negative shocks causing a more pronounced increase in volatility compared to positive shocks of the same magnitude. The NIC derived from the model exhibits a sharp, nonlinear rise in volatility in response to adverse news, closely aligning with empirical patterns observed in financial markets, where declines in asset prices amplify financial risk due to higher debt-equity ratios. The result again highlight the fundamental distinction between econometric and NN models: the former rely on structured restrictions derived from theoretical patterns, while NN models uncover such patterns directly from rich data environments.

\subsubsection{Forecast characteristics of the universal model}
\begin{figure}[h!]
    \centering
    \subfigure[Nvidia (NVDA)]{
        \includegraphics[width=0.6\textwidth]{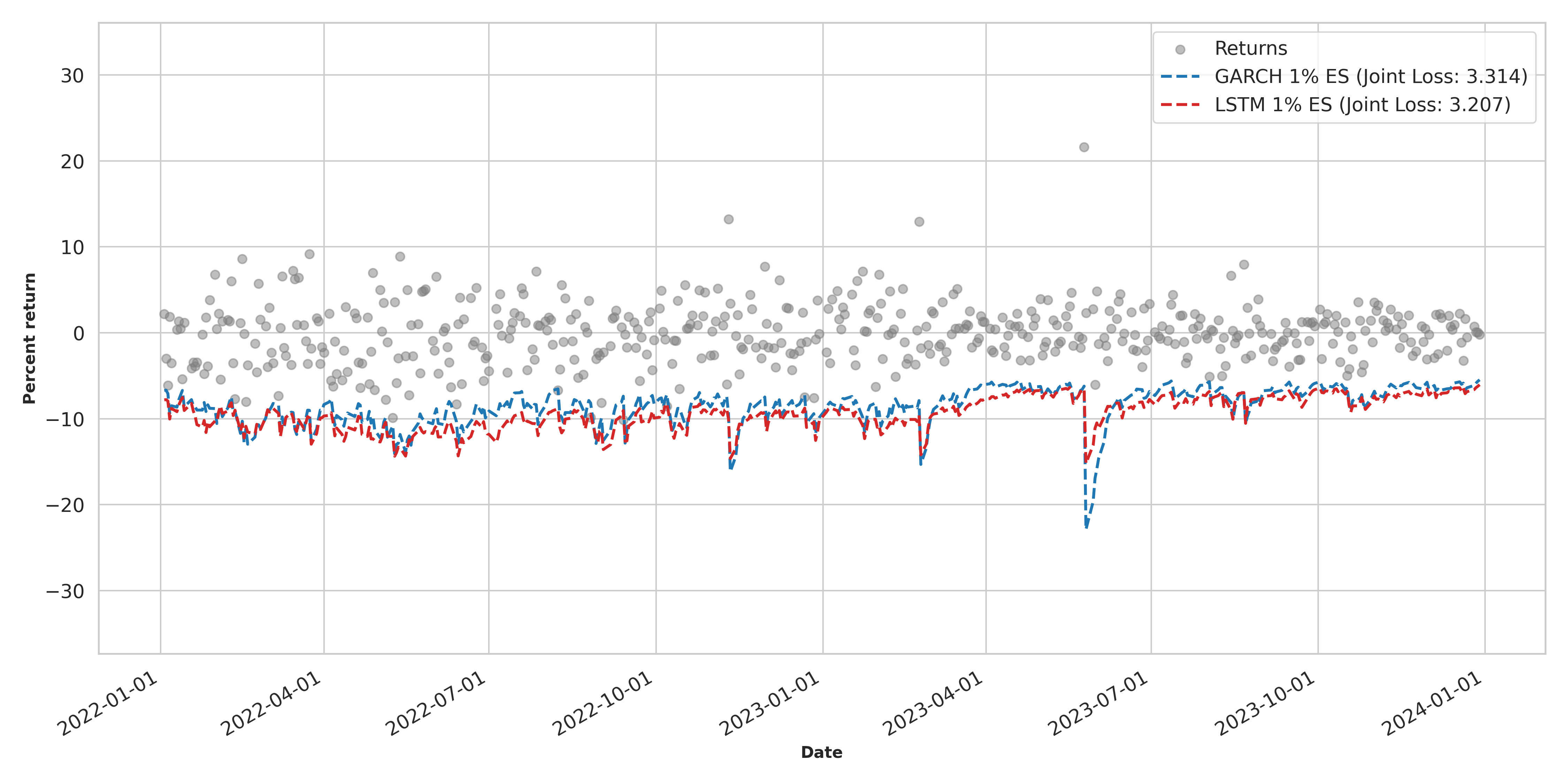}
        \label{fig:es_fc_NVDA}
    }
    \subfigure[Apple (AAPL)]{
        \includegraphics[width=0.6\textwidth]{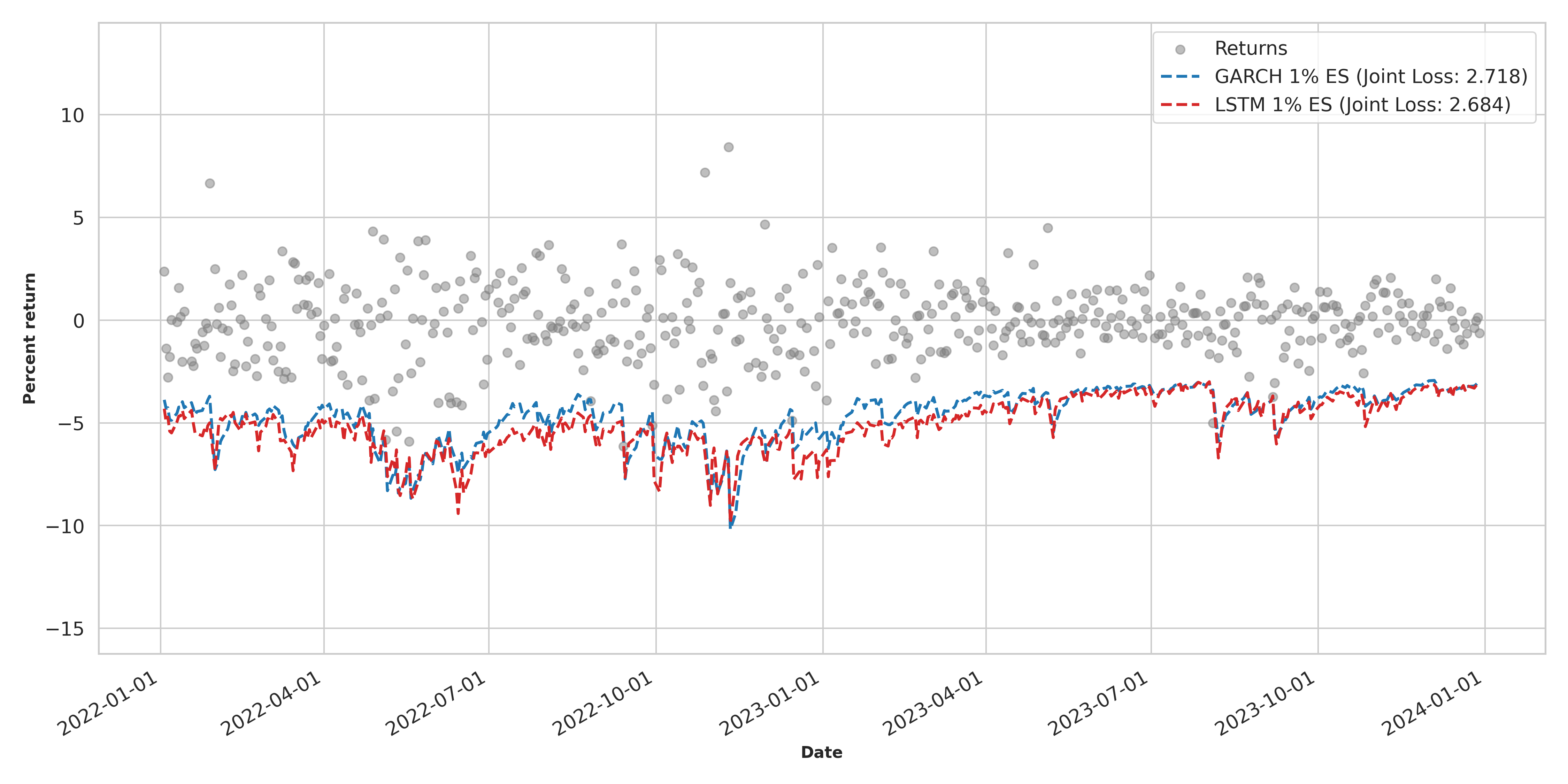}
        \label{fig:es_fc_AAPL}
    }
    \subfigure[Microsoft (MSFT)]{
        \includegraphics[width=0.6\textwidth]{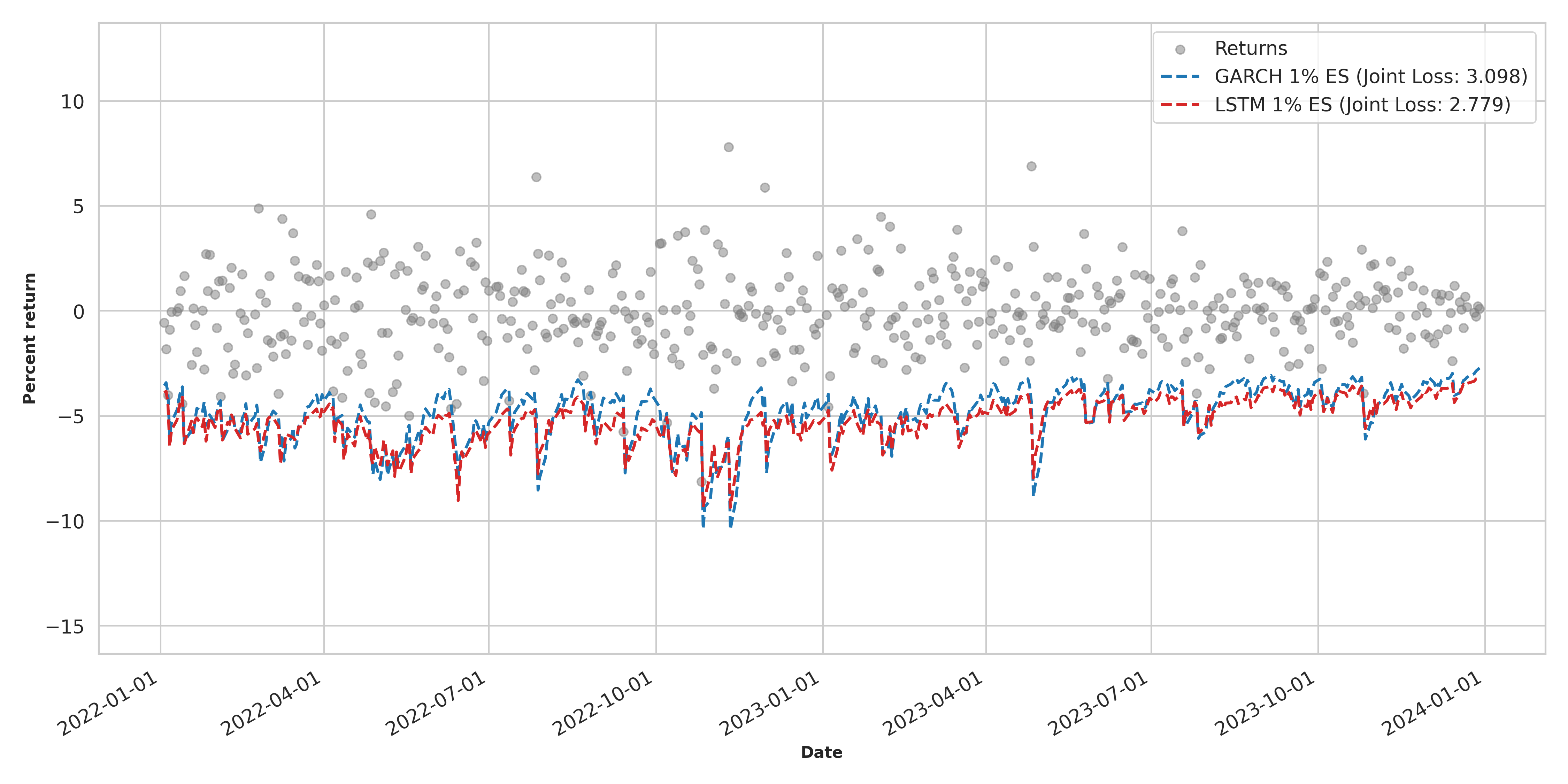}
        \label{fig:es_fc_MSFT}
    }
    \caption{1\% ES forecasts for top 3 market-cap companies.}
    \label{fig:es_fc}
\end{figure}

We now closely examine the forecasts made by the universal model for several individual stocks, comparing their characteristics to those of the forecasts produced by econometric models. 
Figure \ref{fig:es_fc} compares the 1\% ES forecasts generated by the GARCH model and the universal LSTM model for the top three market-cap companies—Nvidia, Apple, and Microsoft—during the out-of-sample period, along with their respective joint losses for the ES forecasts. The results demonstrate that the universal model achieves improved joint loss compared to local GARCH models, while exhibiting markedly different forecast characteristics. The universal model generally produces more conservative forecasts, i.e., more negative ES values, during relatively stable periods when stock returns experience minimal sudden jumps (e.g., for Apple from 01-01-2023 to 01-04-2023). In contrast, during periods of volatile market conditions with sudden jumps in returns, the universal model produces forecasts that are much less extreme and recovers from outliers much faster (e.g., the high shock in Nvidia stock around 2023-06).

\begin{figure}[h!]
    \centering
    \subfigure[Without outliers]{
        \includegraphics[width=0.47\textwidth]{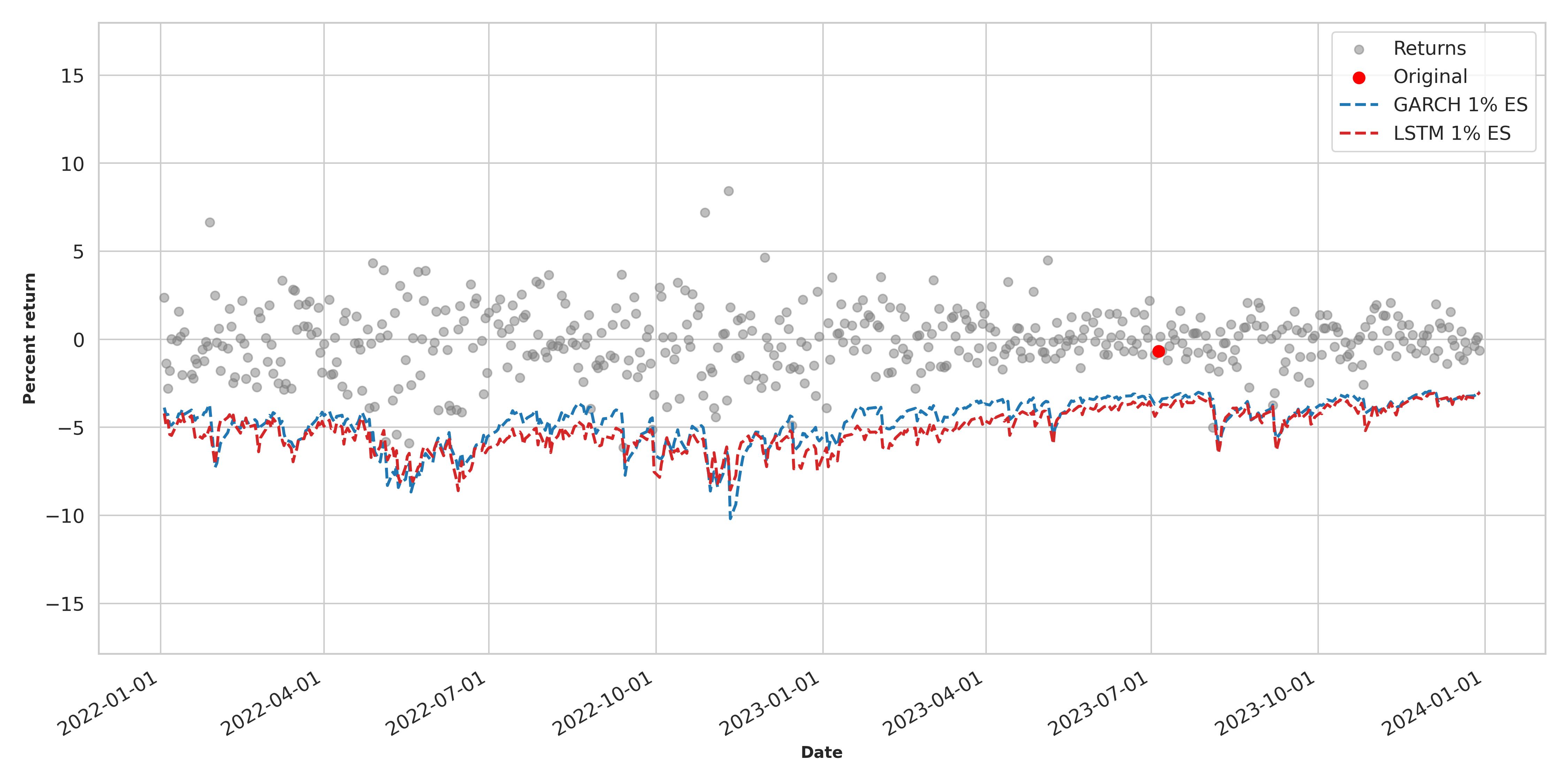}
        \label{fig:es_fc_AAPL_without_outliers}
    }
    \subfigure[With outliers]{
        \includegraphics[width=0.47\textwidth]{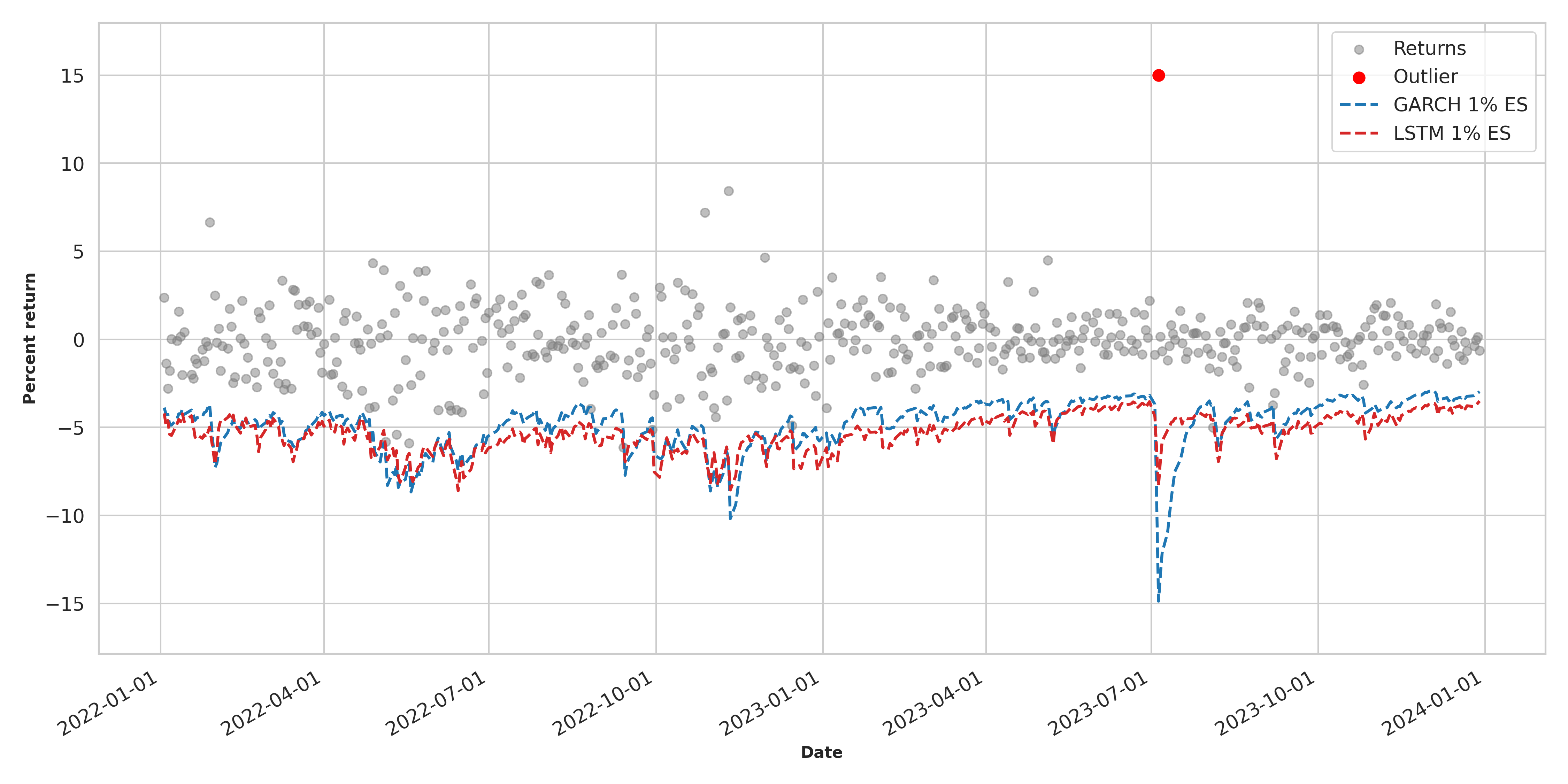}
        \label{fig:es_fc_AAPL_with_outliers}
    }
    \caption{Impact of the outlier on the ES forecasts for Apple stock.}
    \label{fig:es_outliers}
\end{figure}

To further illustrate this difference, we artificially added a single high shock in Apple stock returns and analyzed the reaction of GARCH and universal LSTM models. Figure \ref{fig:es_outliers} plots the ES forecasts during the out-of-sample period, with and without the added outlier. The results show that the universal model responds much more moderately to the outlier, a behavior attributed to its training on a large, diverse dataset of stocks, which significantly enhances its robustness to extreme values. Additionally, the universal model recovers from outliers significantly faster than GARCH models. In this example, the underlying volatility dynamics did not change after the shock. However, the GARCH model generated an extended period of high-volatility forecasts due to the shock, a limitation inherent to the structure of GARCH-family models, where shocks can only have an additive effect. Even if stock returns immediately return to zero after a high shock, the GARCH-family model can only reduce volatility forecasts incrementally at a rate determined by the $\beta$ parameter, leading to slower recovery. In contrast, the universal NN model, unconstrained by such structural limitations, can adjust variance instantaneously while still capturing volatility clustering, allowing them to return to normal volatility forecasts much faster and adapt quickly to rapidly shifting market conditions.

We further examine the performance of the universal LSTM model during high and low volatility years. Table \ref{tab:performance_by_year} presents the percentage improvement in NLL of the universal model relative to GARCH models for each year, along with the average standard deviations of stock returns in that year (higher average standard deviations indicating more volatile years). The results demonstrate that the universal model outperforms the local GARCH models more during periods of pronounced market turbulence. 

\begin{table}[h!]
\centering
\caption{Model performance by year: The middle column reports the percentage improvement in NLL of the universal LSTM model over local GARCH models. The last column shows the average standard deviation of returns for all stocks in each year with higher average standard deviation indicates more volatile years. Both metrics are calculated across all 11,771 stocks. Note: 2020 and 2021 are validation periods, while 2022 and 2023 are testing periods.}
\label{tab:performance_by_year}
\begin{tabular}{lcc}
\toprule
Years & Average NLL Improvement (\%) & Average Std of Returns \\
\midrule
2020 & 2.362 & 3.266 \\
2021 & 1.546 & 2.452 \\
2022 & 1.698 & 2.578 \\
2023 & 1.030 & 2.149 \\
\bottomrule
\end{tabular}
\end{table}

This robustness of the universal model ensures consistent performance across diverse market conditions. By dynamically adapting and recovering quickly from market shocks, the universal model enhances reliability for risk management and decision-making, offering an attractive alternative to econometric models.

We also include a residual analysis in Appendix \ref{sec:residual}, showing that the standardized residuals of the universal model display characteristics consistent with a well-specified GARCH model, including uncorrelated but non-normally distributed series.

\subsection{Portfolio risk forecasts}
Portfolio risk forecasting plays a critical role in the financial industry. This section evaluates the performance of the universal model in forecasting portfolio risk. To this end, we generated 10,000 portfolios using the following procedure:
\begin{itemize}
\item Randomly select the size of the portfolio $M$, where $M \sim \text{Uniform}(10, 50)$.
\item Randomly select $M$ stocks from the pool of 11,771 available stocks.
\item Assign random positive weights summing to 1 to the $M$ selected stocks.
\item Compute the weighted sum of the $M$ stocks to construct an artificial portfolio.
\end{itemize}
The universal LSTM model is then evaluated on these 10,000 portfolios. We note that the universal model is not retrained for these portfolios, i.e. it provides zero-shot forecasts for these portfolios.

\begin{table}[h!]
\centering
\begin{tabular}{lccccc}
\toprule
 & GARCH & GJR & EGARCH & LSTM & Universal LSTM \\
\midrule
NLL & 1.631 & 1.625 & 1.622 & 1.645 & {\bf 1.614} \\
QLoss 1\% & 0.068 & 0.064 & 0.062 & 0.074 & {\bf 0.054} \\
QLoss 2.5\% & 0.128 & 0.123 & 0.121 & 0.133 & {\bf 0.105} \\
JointLoss 1\% & 2.307 & 2.298 & 2.289 & 2.351 & {\bf 2.117} \\
JointLoss 2.5\% & 2.102 & 2.087 & 2.075 & 2.143 & {\bf 1.960} \\
\bottomrule
\end{tabular}
\caption{Risk metrics for the local models and universal model averaged over all 10,000 portfolios. For all metrics, lower values indicate better performance.}
\label{tab:risk_management_portfolio}
\end{table}

\begin{table}[h!]
\centering
\begin{tabular}{lccccc}
\toprule
 & GARCH & GJR & EGARCH & LSTM & Universal LSTM \\
\midrule
NLL & 1647 & 1449 & 1893 & 1748 & {\bf 9151} \\
    & (0.123) & (0.102) & (0.133) & (0.118) & {\bf (0.807)} \\
QLoss 1\% & 1723 & 1521 & 1972 & 1821 & {\bf 9328} \\
          & (0.128) & (0.107) & (0.137) & (0.122) & {\bf (0.819)} \\
QLoss 2.5\% & 1686 & 1492 & 1927 & 1782 & {\bf 9411} \\
            & (0.126) & (0.105) & (0.135) & (0.120) & {\bf (0.832)} \\
JointLoss 1\% & 1746 & 1573 & 1996 & 1847 & {\bf 9233} \\
              & (0.130) & (0.112) & (0.139) & (0.125) & {\bf (0.814)} \\
JointLoss 2.5\% & 1703 & 1534 & 1952 & 1802 & {\bf 9383} \\
                & (0.127) & (0.108) & (0.137) & (0.122) & {\bf (0.825)} \\
\bottomrule
\end{tabular}
\caption{MCS: the number of times each model is included in the set of superior models 5\% significance level and the averaged $p$-value for each model for portfolios. The higher p-values indicate a higher likelihood of inclusion in the SSM.}
\label{tab:risk_management_portfolio_mcs}
\end{table}

Tables \ref{tab:risk_management_portfolio} and \ref{tab:risk_management_portfolio_mcs} report the average one-day-ahead portfolio risk metrics and MCS results, respectively. The findings from the individual stock analysis remain consistent in the portfolio setting. The universal LSTM demonstrates significant improvements in out-of-sample performance compared to the local baselines. Moreover, the global LSTM is almost always included in the MCS for all portfolios—exceeding its forecasting performance for stocks. This superior performance compared to stock volatility forecasts can be attributed to the characteristics of portfolios, which typically align more closely with broader market movements. By leveraging its exposure to a diverse set of stocks during training, the universal LSTM model has learned those broader market patterns extensively resulting in more accurate and robust portfolio volatility forecasts.

\section{Conclusion}
The scale of financial datasets is growing dramatically in modern finance. Yet, the adoption of data-centric methods that leverage the power of NNs and rich data environments remains limited in mainstream econometrics. Through extensive empirical analysis, we demonstrate that one primary reason for this gap is the common practice of treating financial series as heterogeneous and modeling them locally. Our findings show that the global training approach and adequate data size are essential for the success of NN models, and we provide an in-depth analysis of the data scaling effects and attractive features of global NNs that enhance their practicality for real-world financial applications.

Econometric models and NNs represent two fundamentally different modeling paradigms. The former relies on structured restrictions derived from theoretical patterns, whereas the latter uncovers those patterns directly from data. Although each paradigm offers unique strengths, the prevailing local training approach has constrained the potential of NNs in financial time series forecasting. We envision this paper as a starting point for adopting the global training approach as the standard when applying neural networks to financial time series.

\printbibliography

\newpage
\appendix
\renewcommand{\thesection}{A.\arabic{section}}
\renewcommand{\thetable}{A.\arabic{table}}
\renewcommand{\thefigure}{A.\arabic{figure}}
\setcounter{table}{1} 
\setcounter{figure}{1} 

\part*{Appendix A}
The appendix provides supplementary material, including additional results, training configurations, and a glossary of machine learning terminologies.

\section{Gated Recurrent Unit}\label{sec:GRU}
In additional to LSTM, the GRU model of \citet{chung_empirical_2014} has proven efficiency in many machine learning applications. 
GRU uses reset and update gates, denoted $r_t$ and $z_t$ respectively, to regulate the information flow.
It is written as follows
\begin{subequations}
    \begin{gather}
    r_t=\psi\left(W_{r y} y_t+W_{r h} h_{t-1}+b_r\right) \\
    z_t=\psi\left(W_{z y} y_t+W_{z h} h_{t-1}+b_z\right) \\
    \tilde{h}_t=\tanh \left(W_{h y} y_t+W_{h h}\left(r_t \odot h_{t-1}\right)+b_h\right) \\
    h_t=z_t \odot h_{t-1}+\left(1-z_t\right) \odot \tilde{h}_t \\
    \sigma_{t+1}=\operatorname{ReLu}\left(W_{\sigma h} h_t+b_\sigma\right)+1e^{-8}. 
    \end{gather}
\end{subequations}
where $\operatorname{tanh}(\cdot)$ denotes the hyperbolic tangent activation function, $\odot$ denotes the element-wise product.
The matrices $W$ and vectors $b$ are the model parameters. 
\clearpage

\section{Data Summary Statistics}\label{sec:Data}
\begin{table}[h!]
\centering
\caption{Data summary statistics by Exchange.}
\label{tab:data_summary_classical}
\begin{tabular}{lccccc}
\toprule
 & NStocks & Avg Length & Avg Std & Avg Skew & Avg Kurt \\
\midrule
Tokyo Stock Exchange          & 1987 & 2399 & 2.12 &  0.15 & 11.28 \\
Shenzhen Stock Exchange       & 1954 & 2179 & 3.12 &  0.05 &  6.15 \\
Shanghai Stock Exchange       & 1314 & 2165 & 2.81 &  0.01 &  6.48 \\
New York Stock Exchange       & 1243 & 2441 & 2.46 & -0.62 & 23.88 \\
NASDAQ Stock Exchange         & 1123 & 2375 & 2.89 & -0.44 & 23.13 \\
London Stock Exchange         & 1077 & 2206 & 2.22 & -0.46 & 23.73 \\
National Stock Exchange of India & 832 & 2360 & 2.83 & 0.48 &  9.14 \\
Taiwan Stock Exchange         & 743  & 2397 & 2.01 & 0.14 & 10.20 \\
Hong Kong Stock Exchange      & 509  & 2288 & 2.89 & 0.42 & 17.20 \\
Toronto Stock Exchange        & 249  & 2387 & 2.77 & -0.29 & 21.53 \\
Euronext Paris                & 249  & 2499 & 2.09 & -0.09 & 23.01 \\
Frankfurt Stock Exchange      & 198  & 2410 & 3.31 & 0.04 & 17.39 \\
Saudi Stock Exchange          & 149  & 2428 & 2.13 & 0.05 & 10.76 \\
Australian Securities Exchange & 114 & 2206 & 3.04 & -0.28 & 18.82 \\
Johannesburg Stock Exchange   & 30   & 2253 & 2.63 & -0.00 & 32.74 \\
\bottomrule
\end{tabular}
\end{table}

\section{Parameter Estimates of Global GARCH Models}\label{sec:garch_parameters}
\begin{table}[h!]
\centering
\caption{Parameter estimates of global GARCH models by number of training series}
\label{tab:garch_parameters}
\begin{tabular}{rccc}
\toprule
NSeries & $\omega$ & $\alpha$ & $\beta$ \\
\midrule
10 & 0.109 & 0.054 & 0.931 \\
20 & 0.041 & 0.050 & 0.945 \\
40 & 0.051 & 0.053 & 0.943 \\
80 & 0.019 & 0.034 & 0.964 \\
160 & 0.014 & 0.029 & 0.968 \\
320 & 0.015 & 0.030 & 0.969 \\
640 & 0.014 & 0.029 & 0.970 \\
1280 & 0.013 & 0.029 & 0.970 \\
2560 & 0.015 & 0.030 & 0.968 \\
5120 & 0.014 & 0.030 & 0.968 \\
10240 & 0.015 & 0.030 & 0.969 \\
\bottomrule
\end{tabular}
\end{table}

\section{Limit of Data Scaling Effect}
This section discuss the limit of data scaling effect. As Figure \ref{fig:lstm_data_scaling} shows, in our current experiment setting the performance of the global models plateaus when more than 1,280 stock series are pooled. This indicates that beyond a certain threshold, adding additional data from similar sources no longer contributes to performance improvements. This finding aligns with recent research on scaling laws in other areas of neural networks \citep{sorscher_beyond_2022, fernandez_hardware_2024}. To achieve further gains in model accuracy, it becomes necessary to shift focus from data size to data diversity. In other words, incorporating data from varied and distinct sources offers greater potential for improvement than merely increasing the volume of homogeneous data. Incorporating diverse financial instruments, such as derivatives, bonds, and foreign exchange rates, may enable the development of a more robust universal model for financial time series.

\section{Residual Analysis}\label{sec:residual}
\begin{figure}[h!]
    \centering
    \begin{tabular}{cc}
        \subfigure[Standardized residuals (NVDA)]{
            \includegraphics[width=0.48\textwidth]{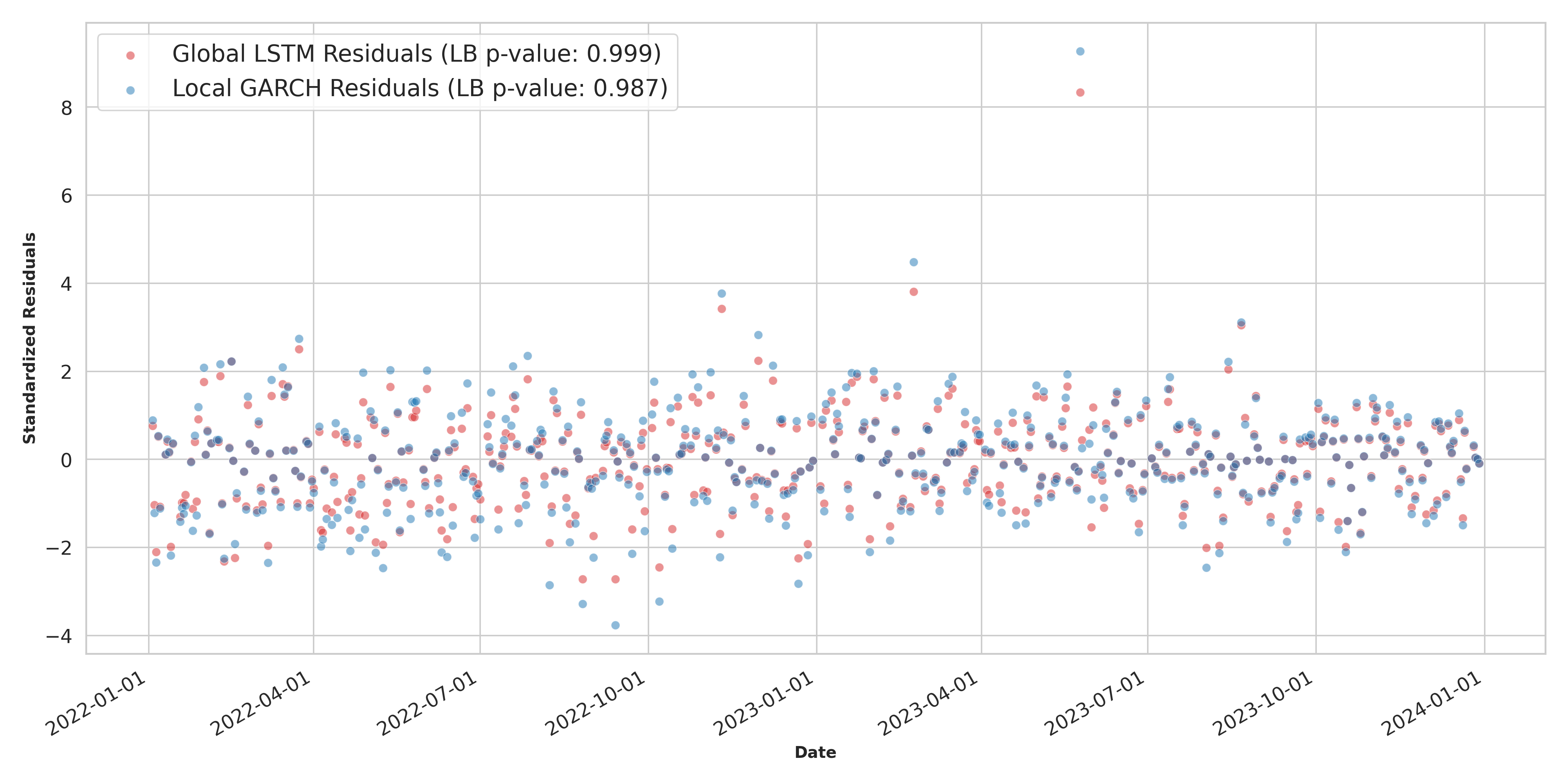}
            \label{fig:standardized_residuals_NVDA_test}
        } &
        \subfigure[Residual distribution (NVDA)]{
            \includegraphics[width=0.48\textwidth]{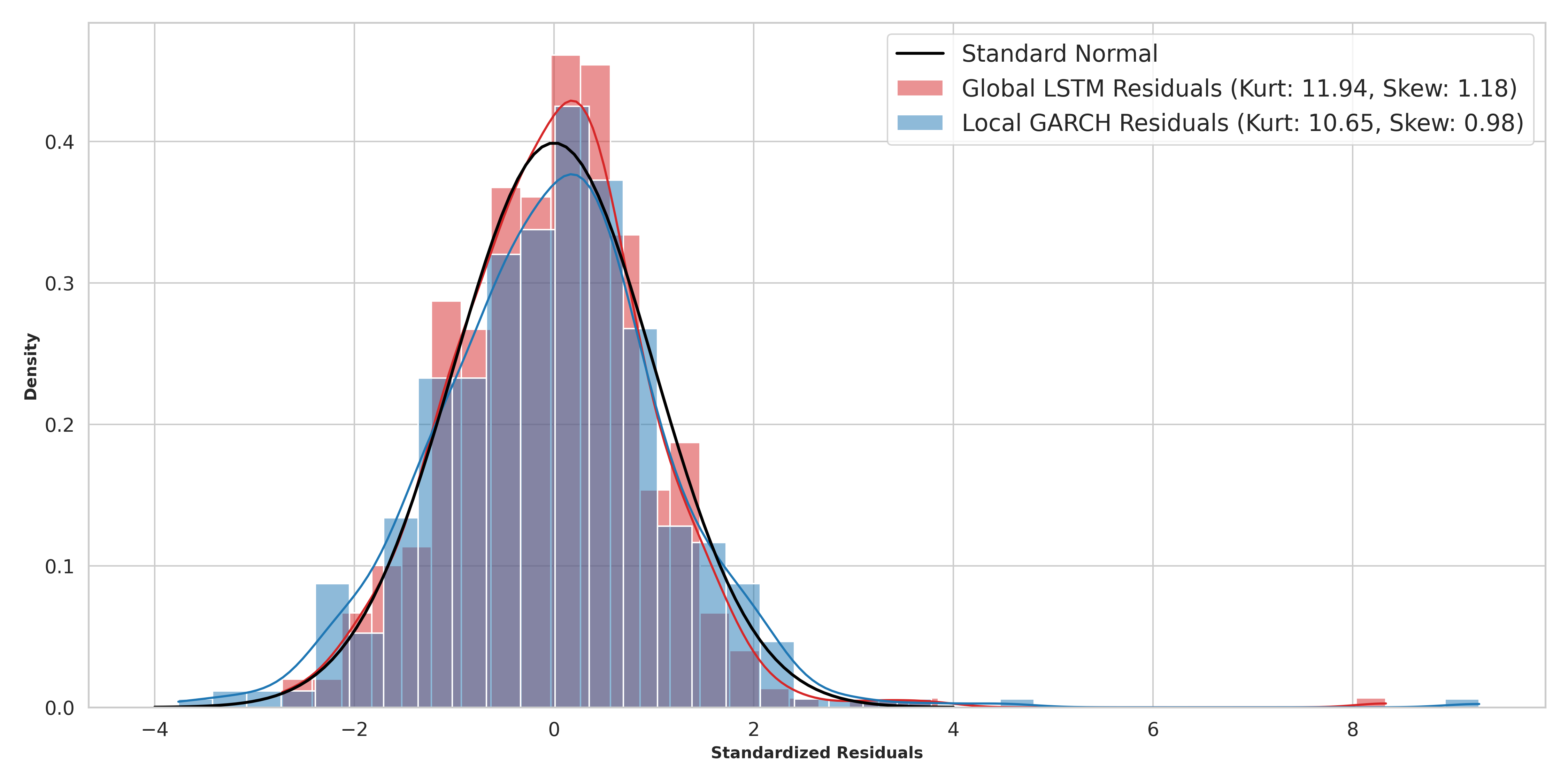}
            \label{fig:residual_distribution_NVDA_test}
        } \\
        \subfigure[Standardized residuals (AAPL)]{
            \includegraphics[width=0.48\textwidth]{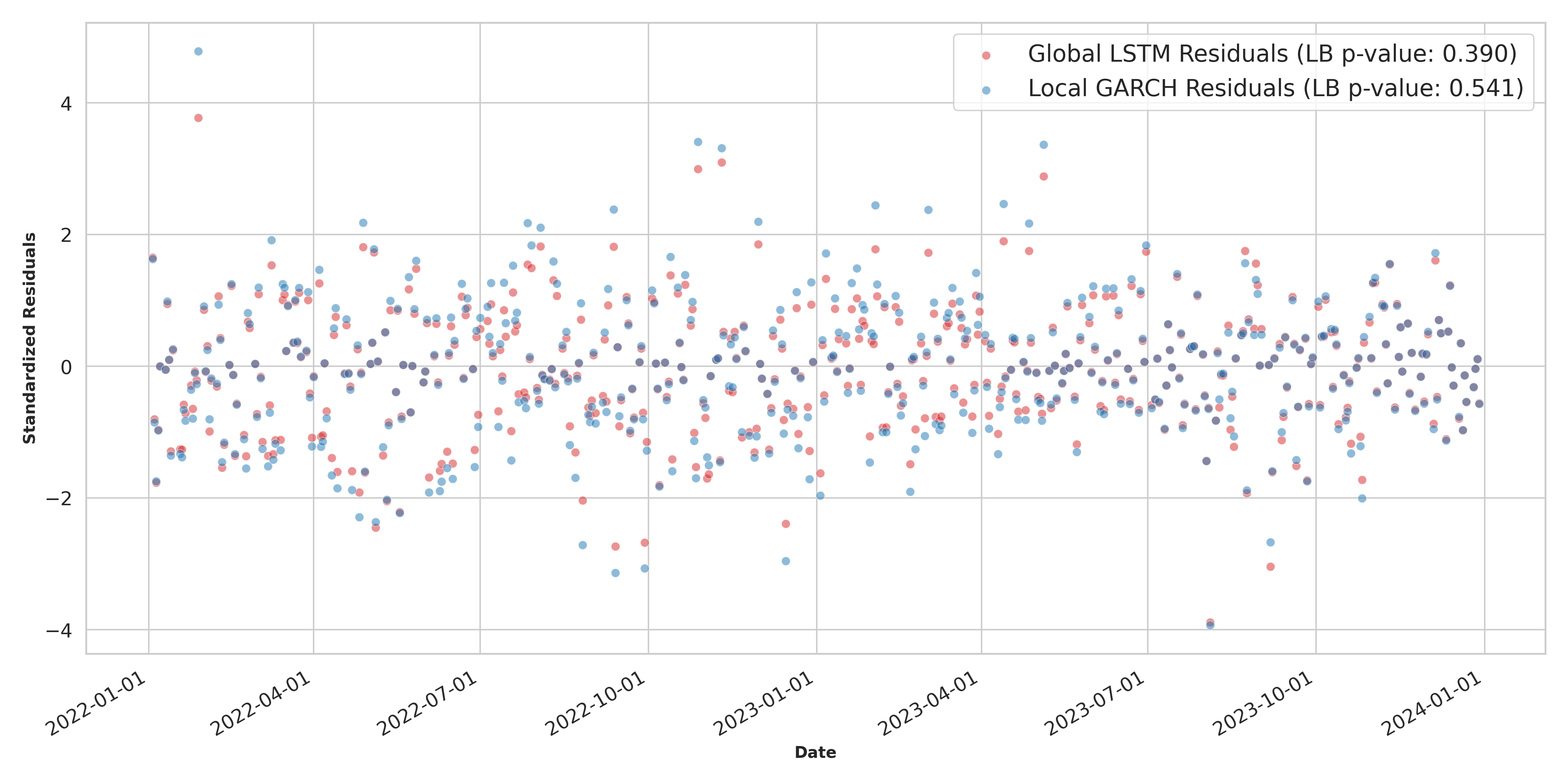}
            \label{fig:standardized_residuals_AAPL_test}
        } &
        \subfigure[Residual distribution (AAPL)]{
            \includegraphics[width=0.48\textwidth]{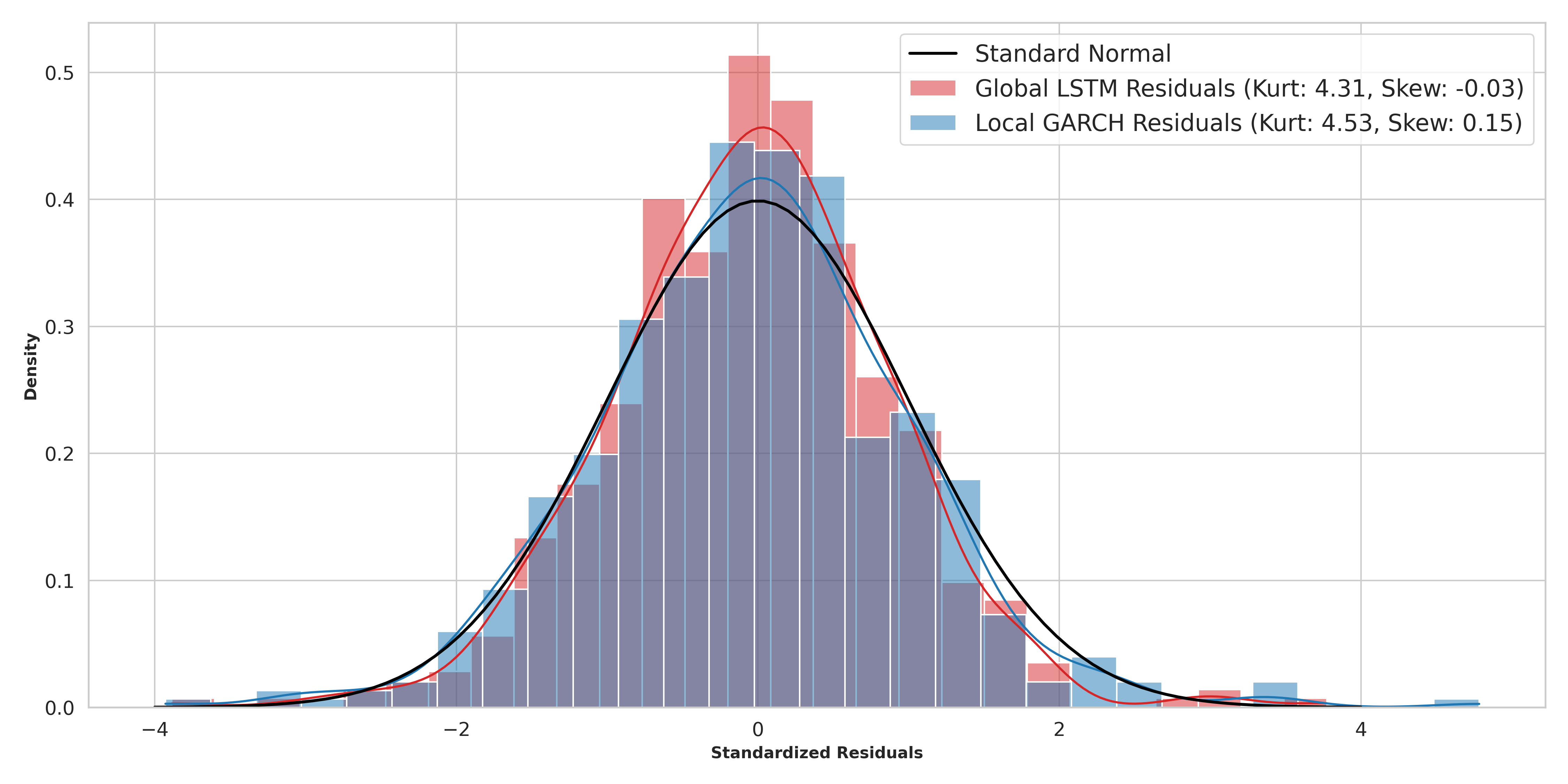}
            \label{fig:residual_distribution_AAPL_test}
        } \\
        \subfigure[Standardized residuals (MSFT)]{
            \includegraphics[width=0.48\textwidth]{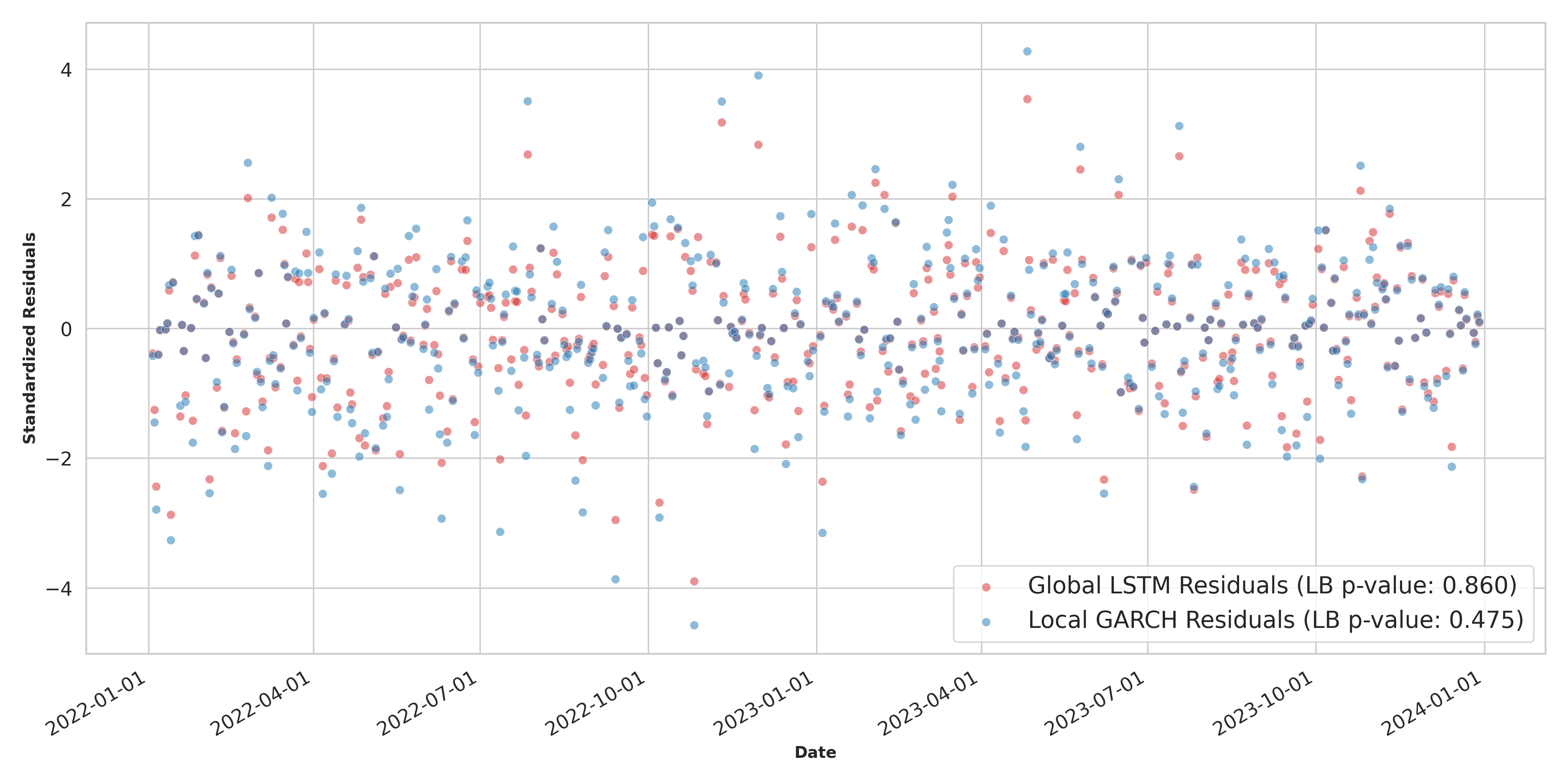}
            \label{fig:standardized_residuals_MSFT_test}
        } &
        \subfigure[Residual distribution (MSFT)]{
            \includegraphics[width=0.48\textwidth]{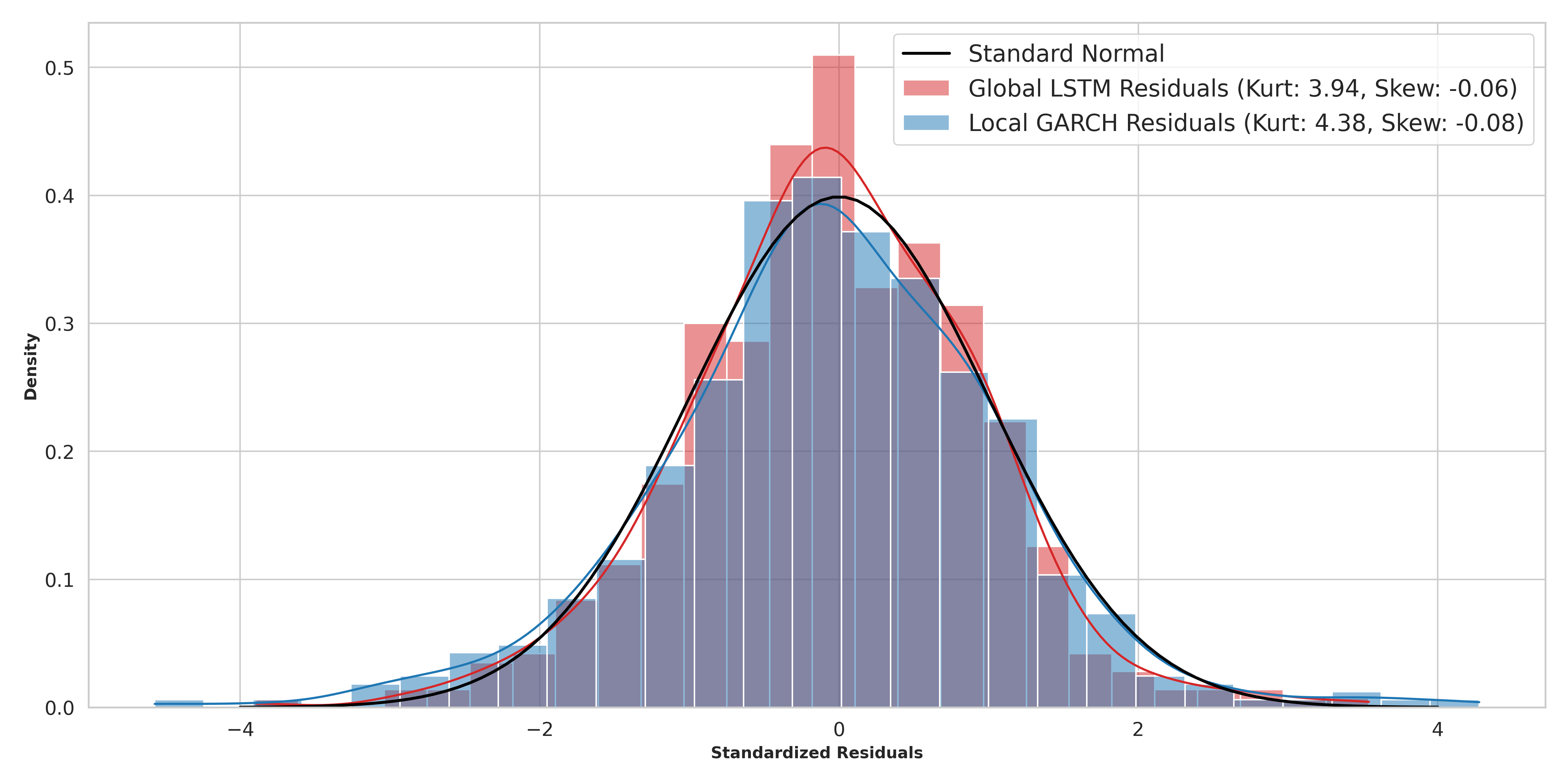}
            \label{fig:residual_distribution_MSFT_test}
        } \\
    \end{tabular}
    \caption{Standardized residuals and their distributions for the test periods of the top 3 companies. The labels on the standardized residual plots display the results of the Ljung-Box (LB) test at lag 10, while the labels on the distribution plots indicate the skewness and kurtosis values.}
    \label{fig:combined_residuals_test}
\end{figure}

This section presents the residual analysis of the universal volatility model. Figure \ref{fig:combined_residuals_test} plots the standardized residuals, including the results of a Ljung-Box (LB) test for autocorrelation in the squared standardized residuals at lag 10, along with their distributions for the top three market-cap companies. The residual analysis highlights the model’s ability to capture key stylized facts in financial time series, particularly the heterogeneity observed in volatility patterns. Similar to GARCH-family models, the universal model effectively accounts for varying degrees of persistence and clustering in volatility across different time frames, reflecting the intricate dynamics of financial markets. The standardized residuals exhibit characteristics consistent with a well-specified model, including uncorrelated but non-normally distributed series.

\section{Training configurations}\label{sec:training_configuration}
The training configuration employed in this paper is designed to ensure consistency across the numerous NN models we trained. Similar to scaling law studies in NLP and CV \citep{kaplan_scaling_2020, zhai_scaling_2022}, we applied nearly the same hyperparameter settings across all NN models.

\paragraph{Optimization}
We optimize the in-sample NLL of all models using the standard Adam optimizer \citep{Kingma:2014vow}. To achieve fast convergence and approach a near-minimum function value, we implement a cosine adaptive learning rate schedule. The learning rate begins at $1 \times 10^{-2}$ and gradually decreases to a minimum of $1 \times 10^{-4}$ over the course of the training. All models are trained for 10000 epochs to ensure sufficient optimization. To prevent overfitting, we apply early stopping, which terminates training if the validation loss does not improve over a specified number of epochs. The patience parameter is set to 1000 epochs, allowing training to continue for up to 1000 additional steps after the last improvement in validation loss. If no improvement is observed during this period, training stops, and the model parameters revert to those corresponding to the lowest validation loss recorded.

All NN training is performed on a consumer-level Nvidia 4090 GPU with an Intel CPU 13900K, using the PyTorch library.  

\paragraph{Mini-batch training and Batch size}
To train global models, we use standard mini-batch training to optimize the objective function \eqref{eq:DeepVol3}. At each iteration, a mini-batch $\mathcal{M}$ of $m$ stock series is randomly selected to provide an unbiased estimate of the objective function:
\begin{equation}
\widehat{\ell}(\mathbf{Y}|\mathbf{\theta}^*) = -\frac{1}{m}\frac{1}{T_{\text{in}}} \sum_{y_{1:T_{\text{in}}}^n\in \mathcal{M}} \sum_{t=1}^{T_{\text{in}}} \log(p(y_t^n | \sigma_t^n)).
\end{equation}
The training process learns the model parameters $\mathbf{\theta}^*$ using the stochastic gradient $\nabla \widehat{\ell}(\mathbf{Y}|\mathbf{\theta}^*)$. For example, with a batch size of $m=10$, each optimization step involves 10 randomly selected stock series. To ensure the same number of optimization steps across models trained with different data sizes, we set the batch size to $\text{Number of Training Stocks}/5$. For instance, a model trained with 10 stocks uses a batch size of 2, while a model trained with 10,240 stocks uses a batch size of 2048.

\paragraph{Handling variable-length stock series}
To leverage modern parallel computing methods, each mini-batch of series must have the same length. This poses a challenge as stock series often vary significantly in length, especially across different exchanges. To address this, we use padding and masking techniques. Padding extends all series within a mini-batch to match the length of the longest series by appending placeholder values (e.g., zeros) to shorter series. Masking is then applied to identify these padded values, ensuring they are ignored during optimization, so that only valid data points are processed by NNs.

\paragraph{Expanding-window forecast and rolling-window forecast}
Our paper uses both expanding-window and rolling-window forecasting methods. In expanding-window forecasting, the model takes the entire return series as input and outputs a volatility series of the same length, where each volatility represents a one-day-ahead forecast of the corresponding return. This method enables the model to leverage all historical returns for making forecasts. In rolling-window forecasting, the model takes a fixed window of return series (e.g., the most recent 252 observations) as input and output only the volatility forecast for the final time step. This approach allows us to evaluate the performance of global NNs with different lengths of historical returns (e.g., the time step importance study in Section \ref{sec:Data scarcity and temporal importance}).

\section{Glossary}
To assist the reader who are not familiar with ML terminologies, we provide a glossary of the terminologies used in the paper, ordered alphabetically.
\paragraph{Activation function} A function that transforms the weighted sum of inputs (from the previous layer) plus a bias term into a nonlinear output. This output serves as the input for the next layer in the network.

\paragraph{Adam} Adam, short for “adaptive moments,” is an optimization algorithm that combines features of RMSprop and momentum methods. It adapts the learning rate for each parameter based on first and second moments of the gradients, enhancing training efficiency.

\paragraph{CV} Computer Vision (CV) refers to the use of machine learning techniques to interpret and analyze visual data, such as images or videos. Common tasks include object detection, image classification, and facial recognition.

\paragraph{Epoch} An epoch refers to a complete pass through the entire training dataset during the optimization process.

\paragraph{Learning rate} The learning rate determines the step size at each optimization iteration, controlling how quickly or slowly the model updates its parameters.

\paragraph{NLP} Natural Language Processing (NLP) involves the application of machine learning models to process and analyze textual data. It includes tasks such as text classification, sentiment analysis, and language translation.

\paragraph{ReLU} The rectified linear unit (ReLU) is an activation function defined as $\text{ReLU}(x)=\max(0, x)$, introducing nonlinearity by outputting zero for negative values and the input value itself for positive values.

\paragraph{Tanh} The hyperbolic tangent (Tanh) is an activation function that transforms inputs to values in the range $(-1, 1)$. It is defined as $\tanh(x) = \frac{\exp(x) - \exp(-x)}{\exp(x) + \exp(-x)}$.

\paragraph{Test set} A subset of the data, typically held out until the final stage of the training process and is used to assess the model’s out-of-sample predictive accuracy.

\paragraph{Training set} A subset of the data used to optimize the model’s parameters.

\paragraph{Validation set} A subset of the data, separate from the training set, used to evaluate the model during training. It provides an unbiased estimate of model performance and is often used for hyperparameter tuning. In our case, as we use the same hyperparameters for all models, validation set is only used for early stopping.

\paragraph{Zero-shot forecasts} The forecasts produced on new time series that is not included in the model training process.
\end{document}